\documentclass[twocolumn,superscriptaddress,nofootinbib]{revtex4-1}

\usepackage{graphicx,xcolor,amsmath}
\usepackage{preview}
\usepackage{hyperref}
\usepackage{bm}
\usepackage{array,mathtools,amssymb,booktabs}
\usepackage{tabularx}
\usepackage{notoccite}
\newcolumntype{C}{>{$}c<{$}}
\AtBeginDocument{
\heavyrulewidth=.08em
\lightrulewidth=.05em
\cmidrulewidth=.03em
\belowrulesep=.65ex
\belowbottomsep=0pt
\aboverulesep=.4ex
\abovetopsep=0pt
\cmidrulesep=\doublerulesep
\cmidrulekern=.5em
\defaultaddspace=.5em
}

\usepackage{subcaption}
\graphicspath{{figs/}}
\definecolor{c1}{rgb}{0.276313, 0.380084, 0.532349}
\definecolor{c2}{rgb}{0.691895, 0.28922, 0.156884}
\definecolor{c3}{rgb}{0.660542, 0.458281, 0.106538}
\definecolor{c4}{rgb}{0.420136, 0.518677, 0.146164}
\definecolor{c5}{rgb}{0.396366, 0.352968, 0.526013}
\definecolor{c6}{rgb}{0.579059, 0.323666, 0.0767903}
\definecolor{c7}{rgb}{0.272924, 0.463876, 0.586762}
\definecolor{c8}{rgb}{0.75, 0.5625, 0.}
\definecolor{mygray}{rgb}{0.5, 0.5, 0.5}

\def\beq{\begin{equation}}
\def\eeq{\end{equation}}

\usepackage{tikz}
\usepackage{tkz-euclide}
\usetkzobj{all}
\usetikzlibrary{backgrounds}
\usetikzlibrary{decorations.text} 
\usetikzlibrary{decorations.pathmorphing}	
\tikzset{
    v/.style={decorate, decoration={snake, segment length=3mm, amplitude=0.75mm}, draw},
    f/.style={draw=black, postaction={decorate},
        decoration={markings,mark=at position .6 with {\arrow[very thick]{latex}}}},
    fb/.style={draw=black, postaction={decorate},
        decoration={markings,mark=at position .4 with {\arrowreversed[very thick]{latex}}}},
    fnar/.style={draw=black},
    g/.style={decorate, draw=black,
        decoration={coil,amplitude=3pt, segment length=3.5pt}},
    s/.style={dashed,draw=black, postaction={decorate},
        decoration={markings,mark=at position .55 with {\arrow[very thick]{latex}}}},
    sb/.style={dashed,draw=black, postaction={decorate},
        decoration={markings,mark=at position .55 with {\arrowreversed[draw=black,very thick]{latex}}}},
    snar/.style={dashed,draw=black,line width =1.25pt},
}

\begin{document}

\title{Pulsar Timing Probes of Primordial Black Holes and Subhalos}

\author{Jeff A. Dror}
\author{Harikrishnan Ramani}
\affiliation{Theory Group, Lawrence Berkeley National Laboratory, Berkeley, CA 94720, USA}
\affiliation{Berkeley Center for Theoretical Physics, University of California, Berkeley, CA 94720, USA}
\author{Tanner Trickle}
\affiliation{Theory Group, Lawrence Berkeley National Laboratory, Berkeley, CA 94720, USA}
\affiliation{Berkeley Center for Theoretical Physics, University of California, Berkeley, CA 94720, USA}
\author{Kathryn M. Zurek}
\affiliation{Theory Group, Lawrence Berkeley National Laboratory, Berkeley, CA 94720, USA}
\affiliation{Berkeley Center for Theoretical Physics, University of California, Berkeley, CA 94720, USA}
\affiliation{Theory Department, CERN, Geneva, Switzerland}

\begin{abstract}
Pulsars act as accurate clocks, sensitive to gravitational redshift and acceleration induced by transiting clumps of matter. We study the sensitivity of pulsar timing arrays (PTAs) to single transiting compact objects, focusing on primordial black holes and compact subhalos in the mass range from $10^{-12}~M _{\odot}  $ to well above $100~M_\odot$.  We find that the Square Kilometer Array can constrain such objects to be a subdominant component of the dark matter over this entire mass range, with sensitivity to a dark matter sub-component reaching the sub-percent level over significant parts of this range.  We also find that PTAs offer an opportunity to probe substantially less dense objects than lensing because of the large effective radius over which such objects can be observed, and we quantify the subhalo concentration parameters which can be constrained.

 \end{abstract}

\maketitle



\section{Introduction}
\label{sec:intro}

Gravity, as the only known coupling of the Dark Matter (DM) with ordinary matter, currently provides the sole window on the nature of the DM. It is via gravitational probes that we have inferred the DM abundance and its behavior as a cold, collisionless fluid.  While these macroscopic properties of dark matter have, so far, given us relatively limited information on the theory of DM, it is still possible, despite the weakness of the gravitational couplings, to unveil much more about its properties through those interactions. Every dark matter model predicts small scale structure inside galaxies, and the type of structure, including their density profile and radius, gives information on the DM's cosmological history, couplings to the visible sector, and to itself. The most popular DM candidate, the Weakly Interacting Massive Particle (WIMP), predicts a power spectrum with weak clustering on small scales, which is difficult to observe experimentally. On the other hand, there are many models that predict an abundance of gravitationally bound structure (which we shall refer to as compact objects) on small scales, and the details of those structures can point to specific models.

There are many models with dynamics in the early universe that give rise to enhanced matter power on small scales.  For example, phase transitions (as for axion (or scalar) miniclusters~\cite{HOGAN1988228,Kolb:1993zz,Kolb:1993hw,Kolb:1994fi,Kolb:1995bu,Zurek:2006sy, Hardy:2016mns,Fairbairn:2017sil,Enander:2017ogx}), early matter domination~\cite{Erickcek:2011us,Barenboim:2013gya,Fan:2014zua,Dror:2017gjq}, and vector bosons produced during inflation \cite{Graham:2015rva}, can all generate enhanced small scale structure. Perhaps the simplest DM substructure consists of primordial black holes (PBH)~\cite{1971MNRAS.152...75H,1974MNRAS.168..399C,1975ApJ...201....1C}, which is (largely) characterized by a single parameter, the PBH mass, $ M $. They arise in many theories with inflationary dynamics producing large density perturbations on small scales, {\em e.g.} \cite{Carr:1994ar,Kawasaki:1997ju,Clesse:2015wea,Orlofsky:2016vbd,Inomata:2017okj}.  They are one extreme of the generic density profile and hence serve as a benchmark for gravitational probes of small scale structure. 
The most direct astrophysical probes of DM substructure are currently derived from a wide variety of lensing experiments. These experiments together constrain monochromatic compact object abundance in the mass range $10^{-10} - 10~M_\odot$ to be a subdominant component of DM. Current searches include lensing of the Large Magellanic Cloud (e.g., MACHO~\cite{Allsman:2000kg}, EROS~\cite{Tisserand:2006zx}, OGLE~\cite{Wyrzykowski2011}), Andromeda ({\em e.g.}, SUBARU~\cite{Niikura:2017zjd}) and stars in the local neighborhood ({\em e.g.}, from KEPLER~\cite{Griest:2011av,Griest:2013aaa}, and Gaia~\cite{VanTilburg:2018ykj}).  Lensing of distant supernovae~\cite{Zumalacarregui:2017qqd} and quasars~\cite{Wilkinson:2001vv} have also been considered. 

There are also a variety of constraints specific to PBHs. From $10^{-13}~M_\odot$ down to $10^{-15}~M_\odot$ (below $10^{-15}~M_\odot$ PBHs evaporate in a time shorter than the age of the Universe), the existence of white dwarfs~\cite{Graham:2015apa} currently constrains PBHs, and femtolensing may do so in the future~\cite{Katz:2018zrn} (see~\cite{1992ApJ...386L...5G,Nemiroff:1995ak} for earlier work).  Above $100~M_\odot$ a variety of constraints from structure formation and Planck fairly severely constrain the PBH abundance, see ~\cite{Carr:2009jm} for a review. There remains a controversial window between $\sim 1-100~M_\odot$ where PBHs \cite{Bird:2016dcv,Clesse2016,Sasaki2018}, and not astrophysical black holes or neutron stars, could give rise to the events in LIGO \cite{Abbott}.  The event rate due to PBHs seems consistent with the LIGO event rate, though it has been pointed out that a myriad of other constraints apply to PBHs in this mass window, including disruption of compact stellar systems such as Eriadnus II~\cite{Brandt:2016aco} and Segue I~\cite{Koushiappas:2017chw}.

From this discussion we see that there are three regimes where current constraints on compact objects are lacking or limited, and hence where any prospect for setting constraints is particularly intriguing.  First, in the mass window between $\sim 1 - 100~M_\odot$ relevant for LIGO signals, where constraints exist but there are substantial astrophysical uncertainties. Second, in the mass window below $10^{-11}~M_\odot$, where lensing constraints currently do not apply. Finally, in the regime where lensing constraints are significant for monochromatic PBHs ($10^{-11} - 10~M_\odot$), but also suffer from astrophysical uncertainties. The constraints in this regime will not have reach to more diffuse subhalos, owing to the requirement that the radius of the object be smaller than the Einstein radius. 

In this paper, we explore an astrophysically clean measurement of DM compact objects via pulsar timing measurements across the entire mass window $10^{-12}-100~M_\odot$, by combining several different gravitational redshift and timing effects in measurements of pulsar periods. Pulsars with millisecond periods, observed over time scales of decades, are known to be remarkably stable clocks. While their periods fluctuate over short times, these fluctuations do not substantially {\em accumulate}. In practice one can define a pulse phase of the signal,
\begin{equation} 
 \phi ( t ) = \phi _0 + \nu \,t + \frac{1}{2} \dot{\nu} \,t ^2 + \frac{1}{ 6} \ddot{\nu} \,t ^3 + ...
 \label{eq:powerseries}
\end{equation}  
where $ \nu $ is the frequency and $ \dot{\nu} $, $ \ddot{\nu} $ are its first and second derivatives. 
The most stable pulsars have frequencies of $ {\cal O} ( {\rm kHz}  ) $ and a spin-down rate of the pulsar, $ \dot{\nu} / \nu   $, ranging from roughly $ 10 ^{ -  23} - 10 ^{ - 20} \text{ Hz}  $, both of which can be fit from the data. Empirically, it is found that $ \ddot{\nu } /\nu $ can be below $  10 ^{ - 31} \text{ Hz}^{2}$~\cite{Liu:2018lmk} and is typically not included in fits to the data, allowing one to place upper bounds on processes that would produce a non-negligible $\ddot{\nu }$. Furthermore, any process which induces a modification of the phase, 
\begin{equation} 
 \delta \phi \equiv \int d t \, \delta \nu ( t ) \label{eq:dphase}
\end{equation} 
can be constrained using pulsar timing measurements.

The quality of pulsar timing data is determined by three parameters. The first parameter is the root-mean-square (RMS) timing residual, $t_\text{RMS}$. This is determined after finding the frequency, $\nu^\text{fit}$, and its derivative, $\dot{\nu}^\text{fit}$, which minimizes the residual between the timing data, $t_n^\text{data}$, and the timing model, $t_n$, where $t_n$ is found via the relation $2 \pi n = \phi(t_n)$ from Eq.~\eqref{eq:powerseries}. This gives
\begin{equation}
t_{\rm RMS} \equiv \sqrt{ \frac{1}{N}\sum_n (t^\text{data}_n - t^\text{fit}_n )^2 },
\end{equation}
where $N$ is the number of data points, and $t_n^\text{fit}$ is $t_n$ with $\nu = \nu^\text{fit}$, $\dot \nu = \dot{\nu}^\text{fit}$ and all higher order terms dropped. The minimized residual is typically $t_\text{RMS} \sim \mu\text{sec}$. The other two parameters are the observation time of the pulsar, $ T \sim 10$ years, and the time between measurements, $\Delta t \sim 2$ weeks (also known as the cadence). Clearly the pulsars with the most power to constrain substructure are those with smaller RMS noise, longer observation times, and shorter cadence.

Pulsar timing data can probe DM compact objects since a transit near the timing system will give rise to a change in the observed frequency of the pulsar. We consider changes in the observed frequency of the pulsar due to two effects. First, there can be a gravitational time delay due to a changing gravitational potential affecting the photon geodesic as it moves along the line of sight -- this is known as a {\em Shapiro} time delay, and was proposed as a probe of dark matter in~\cite{Siegel:2007fz}. Second, the presence of compact objects can lead to an acceleration of the Earth or pulsar, also changing the observed pulsar period -- this is the {\em Doppler} effect, and was proposed as a signal of dark matter in~\cite{Seto:2007kj}. These accelerations are optimal for studying smaller masses and are typically more sensitive than Shapiro delays, though in some parameter space, as we will explore in detail, Shapiro delays can be more sensitive due to the long baseline.
 
The signal from a transiting compact object will look different depending on the relevant timescale, $ \tau $, associated with the motion of the compact objects (here we use this variable schematically but give it an explicit, mass-dependent meaning in later sections). If we denote the observation time of a pulsar as $ T $, then \textit{dynamic} signals correspond to $ \tau \ll T $, and will appear as blips in the pulsar timing data (analogous to glitches which have been observed in millisecond pulsar data~\cite{Cognard:2004av,McKee:2016qtx}). \textit{Static} signals, with $ \tau \gg T $, will not be observable as blips but instead as a non-negligible contribution to the second derivative of the frequency, $ \ddot{\nu } $. 

The idea of using pulsar timing to probe dark matter substructure has a long history. The static contribution of the Shapiro time delay was suggested as a probe of PBHs in~\cite{Clark:2015sha,Schutz:2016khr}, while searches for dynamic signals were considered for single events in~\cite{Siegel:2007fz,Seto:2007kj,2012MNRAS.426.1369K}, and multiple events in \cite{Baghram:2011is}.  None of these analyses, however, considered how the signals were related to each other in the relevant regime of validity. Our results extend, and differ from, previous results as follows. First, we carry out the first analysis to correctly consider all forms of timing signatures, in the dynamic and static limit, and for both Doppler and Shapiro effects. We comment on the interplay between these four signals and their complementary sensitivity in different mass ranges. The comparative analysis has important implications for signals; for example, in contrast to previous work, we find that the Doppler signal dominates in the static limit, substantially modifying the derived constraint. Second, we perform the first study of the single event `blip' signal shapes and compute these shapes in three dimensions; this extends and improves on the previous limits derived in~\cite{Seto:2007kj,2012MNRAS.426.1369K,Kashiyama:2018gsh}. Third, we perform projections for current and future pulsar timing experiments in all of the signal regimes, correctly incorporating the impact of the measurement cadence on the constraint for the first time. Lastly, we study the impact of the size of compact objects, parameterized in terms of the profile, on the constraints derived. Note that we do not consider a multi-event (or statistical) signal, as studied in~\cite{Baghram:2011is}. While we expect that such an analysis will extend the reach at the low mass end (below ${\cal O} ( 10^{-9} \, M_\odot ) $ for Doppler signals and below $ {\cal O} (10 ^{ - 4} \, M _{\odot} ) $ for Shapiro signals), due to the more complicated nature of the signal, we reserve study for future work \cite{future}. 

The outline of this paper is as follows.  In Sec.~\ref{sec:ShapiroVsDoppler} we describe static and dynamic signatures of transiting compact objects, for both Doppler and Shapiro effects, being careful to delineate the dividing line between the regimes. Next, in Sec.~\ref{sec:methods}, we detail the size of the signals expected in the dynamic and static regimes for both Doppler and Shapiro signals. Then we present the analytic and numerical results in Sec.~\ref{sec:results}, projecting constraints on the fraction of DM in PBHs (or PBH-like subhalos) which can be probed using pulsar timing.  These results are extended to more diffuse subhalos in Sec.~\ref{sec:size}, where we show that PTAs have sensitivity to much more extended objects than lensing searches. Finally, in Sec.~\ref{sec:conclusion}, we summarize our results and suggest ways in which the analysis can be extended.  

\section{Pulsar Timing Signatures from Doppler and Shapiro Effects}

\label{sec:ShapiroVsDoppler}
Transiting compact objects give rise to two different effects in the time of arrival of pulses from pulsars. The first, the Doppler effect, arises from an acceleration of the Earth or the pulsar.  The Shapiro effect, on the other hand, is a gravitational redshift effect along the photon geodesic. Both of these effects cause the photon arrival time to be shifted from the unperturbed propagation value. The constant terms inside of these time shifts are unobservable as they can be absorbed by a redefinition of the unperturbed travel time.  We thus consider time-dependent changes which generate a shift in the pulsar frequency, $\delta \nu$. For the Doppler and Shapiro signals, we have, \footnote{Here we assume a weak field approximation, $\Phi \ll 1$, a slowly varying potential during the interaction time scale ($\Phi(r + v r) \simeq \Phi(r)$), where $r$ is the distance of closest approach, and large orbit eccentricity.}
\begin{align}
\left( \frac{\delta \nu}{\nu} \right)_D &=  \mathbf{\hat{d}} \cdot \int \nabla \Phi \;dt, \label{eq:dopsignalpre} \\
 \left( \frac{\delta \nu}{\nu} \right)_S & = -2\int \mathbf{v} \cdot \nabla \Phi \;dz, \label{eq:shapirosignalpre}
\end{align}
where $\Phi$ is the gravitational potential due to the compact object and $\mathbf{v} $ is its velocity, while $\mathbf{\hat{d}}$ is the direction from the Earth to the pulsar and $z$ parameterizes the path the light takes from the pulsar to the Earth. 
These expressions can be simplified by assuming the compact object is a PBH of mass $M$,
\begin{align}
\left( \frac{\delta \nu}{\nu} \right)_D & =  G M \mathbf{\hat{d}} \cdot\int  \frac{ \mathbf{r}}{r ^3}\,  dt,  \label{eq:dopplermiddle}\\
\left( \frac{\delta \nu}{\nu} \right)_S & = -4 G M \frac{\dot{r} _\times }{r_\times}, \label{eq:shapiromiddle}
\end{align}
where $\mathbf{r}$ is the position of the compact object relative to the pulsar and $\times$ subscript denotes crossing with $\mathbf{\hat{d}}$, $ {\bf r}  _\times \equiv {\bf r} \times {\bf \hat{d} } $. Physically, the Doppler delay derives from integrating over the gravitational field from the compact object and taking the component of the pulsar (Earth) acceleration towards the Earth (pulsar), while the Shapiro delay depends only on components of the position and velocity of the compact object in the direction perpendicular to $\mathbf{\hat{d}}$, as only this gives a time dependent shift to the metric affecting the photons.

As shown in Appendix~\ref{sec:simplifysignal} these expressions can be further simplified to
\begin{align} 
	\left( \frac{\delta \nu}{\nu} \right)_D & =  \frac{GM}{v^2 \tau_D } \frac{ 1 }{ \sqrt{1+ x_D^2}}\;  \left( x_D \mathbf{\hat{b}} - \mathbf{\hat{v}} \right) \cdot \mathbf{\hat{d}} , \label{eq:dopplersignalfinal} \\
	\left( \frac{\delta \nu}{\nu} \right)_S  & =   \frac{4 G M}{\tau_S} \frac{x_S}{1+x_S^2}, \label{eq:shapirosignalfinal}
\end{align}  
where we have taken the motion of the transiting object as ${\bf r} = {\bf r}_0 + {\bf v} t$.  We define $ x_D \equiv ( t -  t _{D,0} ) / \tau_D $, $x_S \equiv (t -  t_{S,0})/\tau_S$ as normalized time variables. Here, the width of each signal is given by $ \tau_D \equiv \left| {\bf r} _0  \times {\mathbf{v}}  \right| / v^2 $ and $\tau_S   \equiv \left| \mathbf{v}_\times \times \mathbf{r}_\times \right| / v_\times^2$. The times for the passing object to reach its point of closest approach are given by $ t _{D,0} \equiv - {\bf r} _0 \cdot {\mathbf{v}}  / v^2 $,  $t_{S,0} \equiv -  \left( \mathbf{v}_\times \cdot \mathbf{r}_\times \right) / v_\times^2$. For the Doppler delay, the vector pointing from the pulsar to the point of closest approach is given by $ {\mathbf{b}}_D \equiv {\bf r} _0 + {\mathbf{v}} t _{D,0} $. For the Shapiro delay the relevant vector points from the line of sight to the point of closest approach, and is given by $\mathbf{b}_S =  \mathbf{\hat{d}} \times  (  \mathbf{r}_\times + \mathbf{v}_\times  t_{S,0} )  $. From here on we will drop the $D, S$ subscripts which will be apparent by context. 

\begin{figure}
 \includegraphics[width=0.45\textwidth]{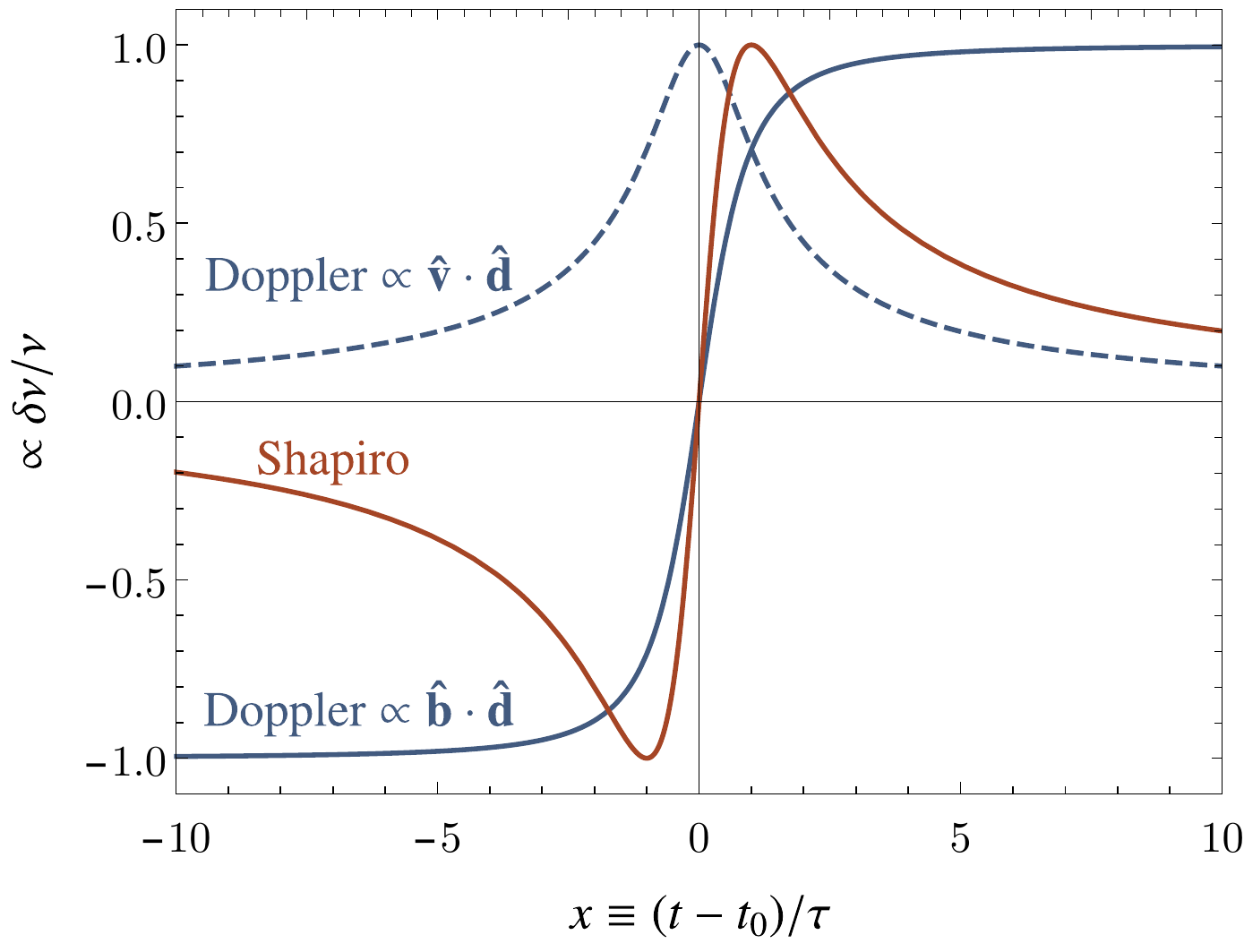}
\caption{Normalized signal shapes observable in pulsar timing data. In general the Doppler signal is a linear combination of the two shapes depending on the object's trajectory, while the Shapiro signal shape is fixed.}
\label{fig:signal}
\end{figure}

The signal shapes are shown in Fig.~\ref{fig:signal}. The Doppler signal has two components depending on the orientation of the incoming object, a transient signal ($ \propto \hat{ {\mathbf{v}} } \cdot \hat{ {\mathbf{d}}  } $) and a non-transient signal ($ \propto \hat{ {\mathbf{b}} } \cdot \hat{ {\mathbf{d}}  } $). The Shapiro signal is always transient regardless of orientation.

Note that one may be tempted to conclude immediately that a Shapiro signal is always subdominant to the Doppler signal, as it is suppressed by $v ^2 $. However, the Shapiro signal is amplified by the long baseline ($ \sim $ kpc) resulting in a much shorter typical timescale, and is able to probe a complementary mass window.  We consider this in detail in the next sections.

\section{Signal Analysis}
\label{sec:methods}
Both the Doppler and Shapiro signals have a characteristic time scale, $\tau$, corresponding to the time for the compact object to pass the line of sight. We note that the signal width ($\tau$) and time to the center of the blip ($ t _0 $) are parametrically the same scale, with the differences being due to the objects' orientation. 

If $\tau \gtrsim T$, where $T$ is the observing time, which we define as the {\em static} limit, we will observe only a small section of the signal, which will have a power series expansion in small $ t /\tau$. As we discussed in Sec.~\ref{sec:intro}, in the static limit, the first two terms in the expansion are unobservable as they are degenerate with the frequency and its derivative. However, a traversing compact object can still be detected by its higher order contributions to Eq.~\eqref{eq:powerseries} corresponding to the coefficient of the ${\cal O} (t^2)$ term (this is degenerate with a measurement of pulsar frequency second derivative, which is known to be small). 

If, on the other hand, $\tau \lesssim  T$ (the {\em dynamic} limit), the whole signal shape is seen and therefore a power series expansion will no longer hold. Since the typical compact object spacing is smaller for lower masses, the distance to the pulsar timing system is also smaller making the dynamic signal dominantly present at lower masses. In this limit, one can look for the entire signal shape in pulsar timing data, analogous to searches for gravitational waves and stellar microlensing events.  Note that deep in this same regime multiple events will typically transit the line of sight over the observation period, where a statistical approach is relevant as proposed in Ref.~\cite{Baghram:2011is}; we leave analysis of such a multi-event signal for future work \cite{future}, where we expect improved reach at lower masses.  

We now compute the observables in pulsar timing experiments for the four different searches (Doppler dynamic and static, Shapiro dynamic and static) and their corresponding signal to noise ratio (SNR). The SNR in each case can be estimated analytically by assuming that the constraints are dominated by the object that comes closest to the pulsar (for the Doppler delay) or the line of sight (for the Shapiro delay). We call this the {\em closest-object} approximation and it holds in most of the parameter space for all four searches.  We also derive an analytic estimate for the split between the dynamic and static limits for the different searches, highlighting their corresponding sensitivity regions.   

In the following sections, we apply the following statistical procedure to determine the reach.  In order to ensure no false positives among the entire pulsar timing array we need to set a threshold for the SNR. In the absence of a signal the SNR at each pulsar is a one-sided Gaussian random variable and therefore we can calculate the $95 \%$ confidence threshold value, $x$, by 
\begin{equation}
\text{Pr}(\text{SNR} < x)^{N_P} = \text{Erf}\left(\frac{x}{\sqrt{2}} \right)^{N_P} =  0.95,
\end{equation}
which gives $x = 3.66$ with $N_P = 200$. We fix the threshold value to be four for simplicity, which ensures no false positives with greater than $ 95 \%$ confidence. In our Monte Carlo simulations, we also require that a signal manifests in $90\%$ of randomly generated universes.  Correspondingly, in our analytic estimates we use the $90$\textsuperscript{th} percentile relevant length scale, denoted by a `min' subscript, derived in Appendix~\ref{sec:mindist}.

\subsection{Static Limit}
\label{sec:static}

In the static regime the constraint is derived from requiring that $ \ddot{ \nu } / \nu $ is small enough to be consistent with the fit shown in Eq.~\eqref{eq:powerseries}. In setting constraints, we assume a dedicated analysis where $\nu$, $\dot \nu$ and $\ddot \nu$ are fit simultaneously (as opposed to the usual procedure which only fits $ \nu $ and $ \dot{\nu} $, and one assumes $ \ddot{\nu} $ is small). This is necessary since otherwise the fits for $ \nu $ and $ \dot{\nu} $ will absorb part of $ \ddot{\nu} $, diminishing the signal. 

Assuming the data can be characterized by white noise and a signal of the form of Eq.~\eqref{eq:powerseries}, the RMS noise can be taken as the uncertainty in each measurement, and the total expected uncertainty obtainable by a least squares fit on the second derivative is found to be~\cite{Liu:2018lmk}\footnote{We disagree with the limit setting procedure employed in~\cite{Schutz:2016khr} which requires the cubic term in the timing residuals to be below $ t _{ {\rm RMS} } $ (corresponding to the condition, $ \ddot{\nu} / \nu \lesssim 6 t _{ {\rm RMS} }/ T ^3 $) since this does not account for the ``sampling factor'' of $ \sqrt{ 2800 \Delta t / T} $. Coincidentally for pulsar timing array data this is an $ {\cal O} ( 1 ) $ correction for most pulsars since $ 10 ^{ 3}  \text{ wk} \sim 10 \text{ yr}$.}
\begin{equation}
\sigma_{\ddot{\nu} / \nu } = 6 \sqrt{\frac{2800 \Delta t}{T}} \frac{t_\text{RMS}}{T^3} \,.
\label{eq:ddotrms}
\end{equation}
Current pulsar data have an uncertainty of ${\cal O} (  10 ^{ - 31}~ {\rm Hz} ^{ 2 } ) $ while sensitivities are projected to reach $ {\cal O} ( 10 ^{ - 33}~ {\rm Hz} ^{ 2 } ) $ for a single pulsar. This allows us to define a suitable SNR,
\begin{equation}
{\rm SNR} \equiv \frac{ | \ddot{\nu} / \nu | }{ \sigma _{ \ddot{\nu} / \nu }}  >4 \,.
\label{eq:SNRstat}
 \end{equation} 

There are several observational challenges in implementing this analysis. First, in addition to DM compact objects there are other sources which produce a contribution to $ \ddot{\nu} $, such as the existence of dark planets near to the pulsar, as well as a genuine spin down of the pulsar~\cite{Liu:2018lmk}\footnote{Objects in our solar system are not an important background since they experience a yearly modulation and are fit for in the analysis.}. Given this, a static search presents challenges as a discovery method, though it can be reliably used to set constraints on the existence of compact objects.  Interestingly, for some mass ranges, compact objects predict a static Doppler signal in conjunction with a dynamic Shapiro signal, discussed later in this section, which would increase the confidence in the measurement and potentially provide some information about the object size (see Sec.~\ref{sec:size}).

We now calculate the expressions for the static contributions of Doppler and Shapiro signal shapes (when the transiting objects are PBHs), and subsequently estimate their contribution to $ \ddot{\nu} $. In the Dynamic subsection we will discuss the division between the static and dynamic signals. 

\subsubsection{Doppler}
To obtain the size of the Doppler signal shape, we take the second time derivative of Eq.~\eqref{eq:dopplersignalfinal} evaluated at $t = 0$, and use the relation between $ \tau $ and $ t _0 $, $ \tau  ^2 + t _0 ^2  = r _0 ^2 / v ^2 $, to find
\begin{align} 
\frac{ \ddot{\nu} }{ \nu } & = \frac{ G M }{ v ^2 \tau } {\bf \hat{d} } \cdot \left[ {\bf \hat{b} } \frac{ 3 v ^5 \tau ^2 }{ r _0 ^5 }+ {\bf \hat{v} } \frac{ 2 r _0 ^2 v ^3 \tau - 3 v ^5 \tau ^3 }{ r _0 ^5 } \right] \,,\\ 
&  = \frac{G M v}{r_0^3}  \left( {\bf \hat{v}}  + 3 \frac{ v t_0 }{  r _0} {\bf \hat{r} _0} \right) \cdot {\bf \hat{d}}\,, 
\label{eq:dnuDstatic}
 \end{align} 
where in the final step we used the relation $ \left| {\mathbf{b}} \right| = \tau v $.

Note that a static signal can never be isolated and will always be due to a collection of compact objects. However, to understand the sensitivity analytically, we can still make progress by employing the closest-object approximation.
A concern in this approach is whether the impact of far away objects is truly small since the number of compact objects at a given distance grows with distance. However, this is usually a small effect for two reasons. First, the signal size has a steep power of $ r _0 $ in the denominator. Second, the contribution to $ \ddot{\nu} $ does not grow coherently with number of objects and the contribution from a single object can be positive or negative (depending on the object's trajectory). Nevertheless, we note that this approach breaks down when the signal is not deep within the static regime, and where the contribution from multiple objects is needed to adequately estimate the signal size.

Therefore to estimate the signal size from Eq.~\ref{eq:dnuDstatic}, we calculate the minimum typical distance of a DM compact object.  This is derived from the minimum distance of randomly distributed points, around $ N_P $ pulsars, to the closest pulsar, and we assume each compact object is of mass $M$. The result is derived in Appendix~\ref{sec:mindist} and we quote the result here,
\begin{align}
	r_{\text{min}} & \simeq   0.8 \left(\frac{M}{N_P f \rho_\text{DM}}\right)^\frac{1}{3} \nonumber \\
	& \sim \frac{ 10^{-2} \text{ pc}}{\left( N_P f \right)^\frac{1}{3}} \left( \frac{M}{10^{-6} \,M_\odot}\right)^\frac{1}{3}.
\label{eq:rminDopstatic}
\end{align}
Roughly speaking, the static signal condition can be taken as $ r _{ {\rm min}}  \gtrsim v  T  $. However, as discussed above, this is the condition that the static analytic estimates are valid, not that a static search cannot be performed, as the static analytic estimates make use of the closest-object approximation; but when many objects are near the static limit boundary, many objects make similar contributions to $\ddot{\nu}$.

With this expression for $  r _{{\rm min}} $ we can now estimate the size of this signal (dropping angular factors),
\begin{equation} 
 \frac{ \ddot{\nu}  }{ \nu } \simeq   \frac{2 G M v}{r_\text{min}^3} \sim 3 \times 10^{-32} \left( \frac{ N_P f }{ 200 } \right)  \text{ Hz}^{2}\,. \label{eq:staticDop} 
\end{equation}  
Notice that the mass dependence has dropped out due to scaling of the minimum distance as $M^\frac{1}{3}$.  We note that single pulsar measurements have $ \sigma _{  \ddot{\nu} / \nu}  \gtrsim  10 ^{ - 31 }~ {\rm Hz} ^{ 2 } $ and so do not have sufficient sensitivity to see this static signal. However, as we will see future arrays profit significantly from an increase in both observation time (as the uncertainty drops as $ \propto T ^{ - 7/ 2} $), and from many more pulsars in the, and so will be capable of measuring such tiny deviations.

In general compact objects will generate a signal if they gravitationally interact with either the Earth or the pulsar, which we label as the `Earth term' and `pulsar term', respectively. Using correlations between pulsars one can reduce the noise that affects only the signals due to a compact object interacting with the Earth. The constraints on a Doppler signal are, however, always dominated by the pulsar term, as opposed to the Earth term. This can be understood in the limit where all pulsars are identical since in that case for the Earth term, $ \sigma _{ \ddot{\nu} / \nu } \propto 1 / \sqrt{ N_P } $, while the signal is constant, such that the SNR scales as $\sqrt{N_P}$. On other hand, for the pulsar term the noise is independent of the number of pulsars, while the signal size grows linearly with $ N_P $, such that the SNR scales as $N_P$. Therefore, for large $N_P$, the pulsar term dominates.

\subsubsection{Shapiro}
The computation of the static Shapiro signal is analogous to the static Doppler signal. Taking the second time derivative of Eq.~\eqref{eq:shapirosignalfinal}, evaluating at $t = 0$, and simplifying with $ t _0 ^2 + \tau ^2 = r _{\times , 0} ^2  / v _\times ^2  $, gives
\begin{equation}
	\frac{\ddot{\nu}}{\nu} = \frac{ 8 G M v _\times  ^3  }{ r _\times  ^3 } \left( \frac{ t _0 v _\times  }{ r _\times  } \right) ^3 \left( 1 - \frac{3 \tau ^2}{t _0 ^2} \right) \,.
\label{eq:dnuSstatic}
\end{equation}
As expected, $\ddot{\nu}$ appears parametrically suppressed compared to the Doppler contribution, though (as commented previously) this is deceiving due to the different, and typically smaller, distance scale in the denominator.  We also find the same power of $ r _0 $ in the denominator, suggesting that we can again use a closest-object approximation up to the boundary between the static and dynamic regimes where multiple objects are crucial for obtaining the correct signal size.  

In a manner similar the Doppler case, we compute the smallest expected distance of a DM compact object to the line of sight toward some pulsar, $r_{\times, \text{min}}$.  This is done in Appendix~\ref{sec:mindist}, with the final result,
\begin{align}
r_{\times, \text{min}}  & \simeq 0.9\sqrt{ \frac{M}{N_P  f \rho_\text{DM} d }}\nonumber \\ 
& \sim \frac{0.2\text{ pc}}{\sqrt{N_P f}} \left( \frac{ M}{M_\odot}\right)^\frac{1}{2} \left( \frac{ \text{kpc}}{d}\right)^\frac{1}{2}.
\label{eq:rminShstatic}
\end{align}
As before, we are now able to estimate the size of the cubic $\ddot \nu/\nu$ in the closest-object approximation. Omitting angular factors, this is
\begin{align} 
\frac{ \ddot{ \nu} }{ \nu} & \simeq   \frac{16 G M v^3}{r_{\times, {\rm min}} ^3} & \nonumber \\ 
& \sim 8 \times 10^{-33} \, \left( \frac{ N_P f }{ 200 } \right)^\frac{3}{2}\left( \frac{M_\odot}{M} \right)^\frac{1}{2} \left( \frac{d}{ \text{kpc}}\right)^\frac{3}{2} \text{ Hz}^{2}. \label{eq:staticSh}
\end{align} 
Note that the mass dependence does not drop out, as it did in the Doppler static case. This can be traced to the geometry involving the distance to the line of sight, which results in the minimum distance scaling as the square root (rather than 1/3 power) of the number density of DM compact objects.  This mass scaling agrees with the analysis of Ref.~\cite{Schutz:2016khr} (though as we commented previously the constraint on $f$ we obtain does not agree because of a difference in the limit setting procedure).  Because the Shapiro static signature is small and subdominant to the Doppler static signal, current pulsar data are unable to constrain a static signal from DM compact objects. Nevertheless in Sec.~\ref{sec:results} we show that future arrays may be able to observe such tiny contributions.  

\subsection{Dynamic limit}
In the dynamic limit a compact object is close enough to the line of sight that it crosses in a time smaller than the observation time. This means that pulsar timing experiments see the entire signal shape, rendering the expansions in the static limit invalid. Specifically, we take the dynamic constraint to be $\tau<t_0< T - \tau $ and we note that this implies $\tau < T/2$. To extract small signals out of a noisy background we use the prescription employed in gravitational wave searches known as the Matched Filter procedure~\cite{Moore:2014lga,Moore:2014eua}. The idea is to take the time-of-arrival data, apply a filtering procedure (namely, we convolute the data with the Weiner filter), and extract an optimal signal to noise ratio (SNR). For simplicity, we work in the limit $ \Delta t \ll \tau, \left| t_0 \right|  \ll T $ such that the measurement is unaffected by cadence or finite width effects (adding these is straightforward but complicates the expressions). Furthermore, we assume that the timing residual noise is white with a variance given by $t_\text{RMS}$, \textit{i.e.}~\cite{Moore:2014lga}
\begin{align}
\langle \delta t(t_1) \delta t(t_2) \rangle & = t_\text{RMS}^2 \Delta t \, \delta(t_1 - t_2) \label{eq:timingresidualvariance} \\
\langle \widetilde{\delta t}(f) \widetilde{\delta t}(f') \rangle & = t_\text{RMS}^2 \Delta t \, \delta(f - f')
\end{align}
The signal we consider here is $\delta \nu / \nu = \dot{\delta t}$, whose power spectrum is given by
\begin{align}
\langle \widetilde{\dot{\delta t}}(f) \widetilde{\dot{\delta t}}(f') \rangle & = (2 \pi)^2 t_\text{RMS}^2 \Delta t \, f^2 \delta(f - f')  \,.\label{eq:newSn}
\end{align} 
Using a one-sided power spectral density for the noise we identify, $ S_{\dot{\delta t}}(f) \equiv  8 \pi^2 t_\text{RMS}^2 \Delta t \, f^2 $, giving a SNR~\cite{Moore:2014lga},
\begin{align}
\text{SNR}^2 = 4 \int_0^\infty   df\frac{|\widetilde{h}(f)|^2}{S_{\dot{\delta t}}(f)} \,, \label{eq:SNR}
\end{align}
where $\widetilde{h}(f)$ is the Fourier transform of the $\delta \nu / \nu$ signal. We now compute this for Doppler and Shapiro signals given in Eqs.~\eqref{eq:dopplersignalfinal} and~\eqref{eq:shapirosignalfinal}.


Unlike the static signal, we expect the backgrounds in the dynamic case to be less worrisome. Most importantly, the characteristic signal shape is unlikely to have significant overlap with other sources of noise. Perhaps the most prominent candidate to mimic a dynamic signal are pulsar glitches, which have recently been observed in millisecond pulsars~\cite{Cognard:2004av,McKee:2016qtx}. However, pulsar glitches are well parameterized by an instantaneous peak in the phase with a subsequent falling exponential.  Thus they have a different frequency structure than the signals of interest here.  A more troubling background is dark baryonic objects. The baryonic mass distribution is, however, peaked near a solar mass, whereas the objects we consider in the dynamic limit have masses $M \lesssim 10^{-2}\, M_\odot$. 

\subsubsection{Doppler}

To find the SNR of the Doppler signal we insert the signal shape of Eq.~\eqref{eq:dopplersignalfinal} into the expression for the SNR in Eq.~\eqref{eq:SNR}.  This is valid because, in contrast to the static case, the fitting procedure for $\nu,~\dot \nu$ is not degenerate with the signal. This gives an SNR of
\begin{align} 
 {\rm SNR}   \simeq \left( \frac{ G M }{ \tau v ^2 }  \sqrt{  \frac{ T ^3 }{ 12 \, t _{ {\rm RMS}} ^2  \Delta t }} \right)  {\bf \hat{b} } \cdot {\bf \hat{d} }   \,. \label{eq:SNRdop}
\end{align} 
Note that we have dropped the term proportional to $ \hat{{\mathbf{v}} } \cdot \hat{ {\mathbf{d}} } $ in Eq.~\ref{eq:dopplersignalfinal}, as it is parametrically suppressed by $\tau/T$, which is small in the dynamic limit.  

The signal size, as well as the transition between the dynamic and static regimes, can be understood by  employing the closest-object approximation, as discussed previously. Inspection of Eq.~\eqref{eq:SNRdop} suggests that a closest object approximation should hold due to the $ 1/ \tau \propto 1/ r _0 $ in the denominator (and the small spread in the velocity distribution), and we have checked this using a Monte Carlo simulation, which we discuss in Sec.~\ref{sec:results}. In order to obtain an estimate of the SNR, we compute an estimate of the minimum $\tau = \tau_\text{min}$, generated by a random set of points. Note that $\tau$ also corresponds to the minimum impact parameter since, $ \left| {\mathbf{b}} \right|  = \tau / v $. We calculate this explicitly in Appendix~\ref{sec:mindist} and quote the result here, 
\begin{align}
 \tau  _{ {\rm min}} & \simeq  \frac{1 }{ v } \sqrt{  \frac{M}{N_P f \rho_\text{DM} v T}} \nonumber \\ 
 & \sim \frac{20 ~ {\rm yr}}{\sqrt{N_P f}} \left( \frac{M}{10^{-9} \, M_\odot}\right)^\frac{1}{2} \left(\frac{20 \text{ yr}}{T} \right)^\frac{1}{2}.
\label{eq:rminDopdynamic}
\end{align}
Combining Eqs.~\eqref{eq:SNRdop} and \eqref{eq:rminDopdynamic}  gives a good estimate of the largest SNR for a given mass and DM density. We can further estimate the condition for the nearest object to be in the dynamic limit, meaning $ \tau  _{ {\rm min}} \lesssim T/2   $,
\begin{equation} 
M \lesssim 4 \times 10 ^{ - 8}~M _{\odot} \left( \frac{N_P f}{ 200 } \right) \left( \frac{ T }{ 20~{\rm yr} } \right) ^3  
\label{eq:MStatDynDopp}
\end{equation} 
For such small masses, these signals are sufficiently quick to appear as a transient in pulsar timing experiments, while for larger masses the Doppler signal can only appear in a static search which was detailed previously.

Again one can compare contributions from the pulsar and Earth terms assuming the pulsars are identical. For dynamic signals, the Earth term scales as $S _n \propto 1 / N_P$, while the signal is constant. However, for the pulsar term the noise is independent of the number of pulsars, while the signal size grows linearly with $ N_P $. Therefore, the SNR scales identically with $ N_P $ for the pulsar and Earth term.\footnote{This is in contrast to the results presented in~\cite{2012MNRAS.426.1369K}, which claims to achieve more powerful constraints with the Earth term at lower masses. The discrepancy can be traced to parametrically different estimates of the impact parameter, $b$. Ref.~\cite{2012MNRAS.426.1369K} assumes $b \sim v T$, based on dimensional analysis, whereas we derive more specific estimates.} Therefore, deep in the dynamic limit their constraints should be comparable. However, for the pulsar term the dynamic condition, $\tau < T/2$, is easier to satisfy since $\tau \propto 1/\sqrt{N_P}$, whereas for the Earth term $\tau$ is independent of $N_P$. Thus for larger masses we expect the pulsar term to become more sensitive. For simplicity we only use the pulsar term but note that one could achieve improved sensitivity deep in the dynamic limit by studying both of these contributions.

\subsubsection{Shapiro}
Finally we arrive at the dynamic Shapiro signal. To find the SNR of the dynamic Shapiro signal we insert the signal shape of Eq.~\eqref{eq:shapirosignalfinal} into the expression for the SNR in Eq.~\eqref{eq:SNR} as described in Appendix~\ref{sec:simplifysignal}. This gives an SNR of 
\begin{align} 
 {\rm SNR}    & =  G M \sqrt{  \frac{ 32 \, T }{ t _{ {\rm RMS}} ^2 \Delta t } }\,. \label{eq:SNRshap}
\end{align} 
The SNR is independent of $\tau$ in the $\tau \ll T$ limit. This can be understood intuitively. For a transient signal, the SNR scales linearly with the width of the signal however the signal size scales as the $ 1 / r $-potential and hence $ 1 / \tau $, conspiring to produce an SNR independent of $ \tau $ at leading order.   
Nevertheless, keeping higher order terms in the SNR leads to an expression that eventually decreases as $\tau$ approaches $T$ from below. 

Since far objects can produce a signal, the closest object approximation breaks down more quickly than for the dynamic Doppler signal, and one should account for the multiple events in the SNR, reserved for future work~\cite{future} with the methods proposed in~\cite{Baghram:2011is}. Accounting for multiple events necessarily involves a random signal shape and therefore the matched filter procedure used here will not be applicable. However, even with a different signal analysis technique an SNR accounting for the multiple events is expected to be larger than this SNR, such that the constraints quoted here are conservative.
With a single blip analysis this independence results in a minimum mass at which no signal will be seen for any given pulsar, which will later result in a hard cutoff in the projected constraints.

As before, the minimum $ \tau $ is related to the minimum impact parameter of a set of randomly generated points near an infinite line, $ \tau = \left| {\mathbf{b}} \right| / v $. Thus we obtain a minimum signal width,   
\begin{align}
 \tau _\text{min}  & \simeq  \frac{2}{ v }\frac{ M}{N_P  f \rho_\text{DM} v T d}\,, \nonumber \\ 
	& \sim \frac{20~{\rm yr}}{N_P f} \left( \frac{M}{10^{-4} \, M_\odot}\right) \left( \frac{20 \text{ yr} }{ T } \right) \left(  \frac{ \text{kpc}}{ d} \right)\,.
\label{eq:rminShdynamic}
\end{align}
We can now use this to estimate the maximum mass which will generate a dynamic Shapiro signal, $\tau _{ {\rm min}} \lesssim T/2 $:
\begin{equation} 
 M \lesssim  10 ^{ - 2 }~M _{\odot} \left( \frac{ N_P f }{ 200} \right) \left( \frac{ T }{ 20~ {\rm yr} } \right) ^2 \left( \frac{ d }{ {\rm kpc}} \right). 
 \label{eq:MStatDynShap}
\end{equation} 


\section{Constraints on Primordial Black Holes}
\label{sec:results}
We are now prepared to compute the sensitivity of PTAs to PBHs using the signatures we have discussed. Assuming a null result, we study the capability of the searches to set constraints on DM compact objects in the $ ( M ,f ) $ plane, where $f \equiv \Omega  / \Omega _{ {\rm DM}} $ denotes the fraction of dark matter contained in PBHs of mass $ M $. 

Before setting constraints let us briefly comment on current and future pulsar timing capabilities. To estimate the capabilities of current pulsars we compiled data from PPTA~\cite{Reardon:2015kba}, EPTA~\cite{Desvignes:2016yex}, Nanograv~\cite{Arzoumanian:2017puf}, as well as the combined international collaboration, IPTA~\cite{Verbiest:2016vem}, culminating in 73 unique pulsars (for 13 of these pulsars no distance was quoted, so we assume a distance typical of the rest of the set, 1 kpc from the Earth). For the 2015 data releases we assume an additional three years' observing time in order to derive constraints corresponding to current data.
In setting our limits with current data we use parameters from this set without any additional approximations (in particular, we do not resort to approximating the current data as an identical set of pulsars with some chosen parameters). Since pulsar timing precision improves quickly with observation time, continuing to observe these pulsars results in a rapid improvement in the ability of PTAs to discover DM compact objects. Nevertheless, with the upcoming construction of the Square Kilometer Array (SKA)~\cite{Keane:2014vja} (and its already running precursor, MeerKAT~\cite{Bailes:2018azh}), the number of high precision millisecond pulsars is expected to dramatically increase with the potential of uncovering every millisecond pulsar beaming toward Earth in the entire Milky Way. The particular capabilities of the future millisecond pulsar set are not well-known, due to uncertainties both in final capabilities of the array and the number of detectable pulsars in our galaxy. In forming our projections we use the Phase II numbers in Ref.~\cite{Rosado:2015epa} which correspond to 200 millisecond pulsars with 50 ns timing, and two week cadence. Furthermore, we take the typical distance of a pulsar from the Earth to be five kpc. We also present results for a more optimistic case, where SKA finds 1000 millisecond pulsars, with 25 ns timing, which can be observed with weekly cadence and have a typical distance from the Earth of ten kpc. The assumed experimental parameters are summarized in Table~\ref{table:one}.

\begin{table}
    \begin{tabular}{lr@{\hskip 5\tabcolsep}r@{\hskip 5\tabcolsep}r@{\hskip 5\tabcolsep}r@{\hskip 5\tabcolsep}r@{\hskip 2\tabcolsep}}   
 \toprule[0.15em]
         & $T$  [yr]  & $t_{\rm{RMS}}$ [ns] & $\Delta t$ [wk] & $d$ [kpc] & $N_P$  \\\midrule[0.075em] \addlinespace[0.5em]
         Current  & $5-30$ & $50 - 10 ^{ 4 }  $  & $ 1 - 4 $  & $ 0.5-5 $  & 73 \\
         SKA & 20 & 50 & 2 & 5 & 200\\
         Optimistic  & 20 & 25 & 1 & 10 & 1000 \\ 
\bottomrule[0.15em]
    \end{tabular} 
    \caption{Summary of timing parameters that characterize pulsar timing capabilities. SKA projections are taken from~\cite{Rosado:2015epa}. Current constraints are compiled from various sources~\cite{Desvignes:2016yex,2016MNRAS.458.1267V,Arzoumanian:2017puf,Reardon:2015kba} as described in the text.}
    \label{table:one}
\end{table}

 \begin{figure*}[ht!]
   \includegraphics[width=.75\textwidth]{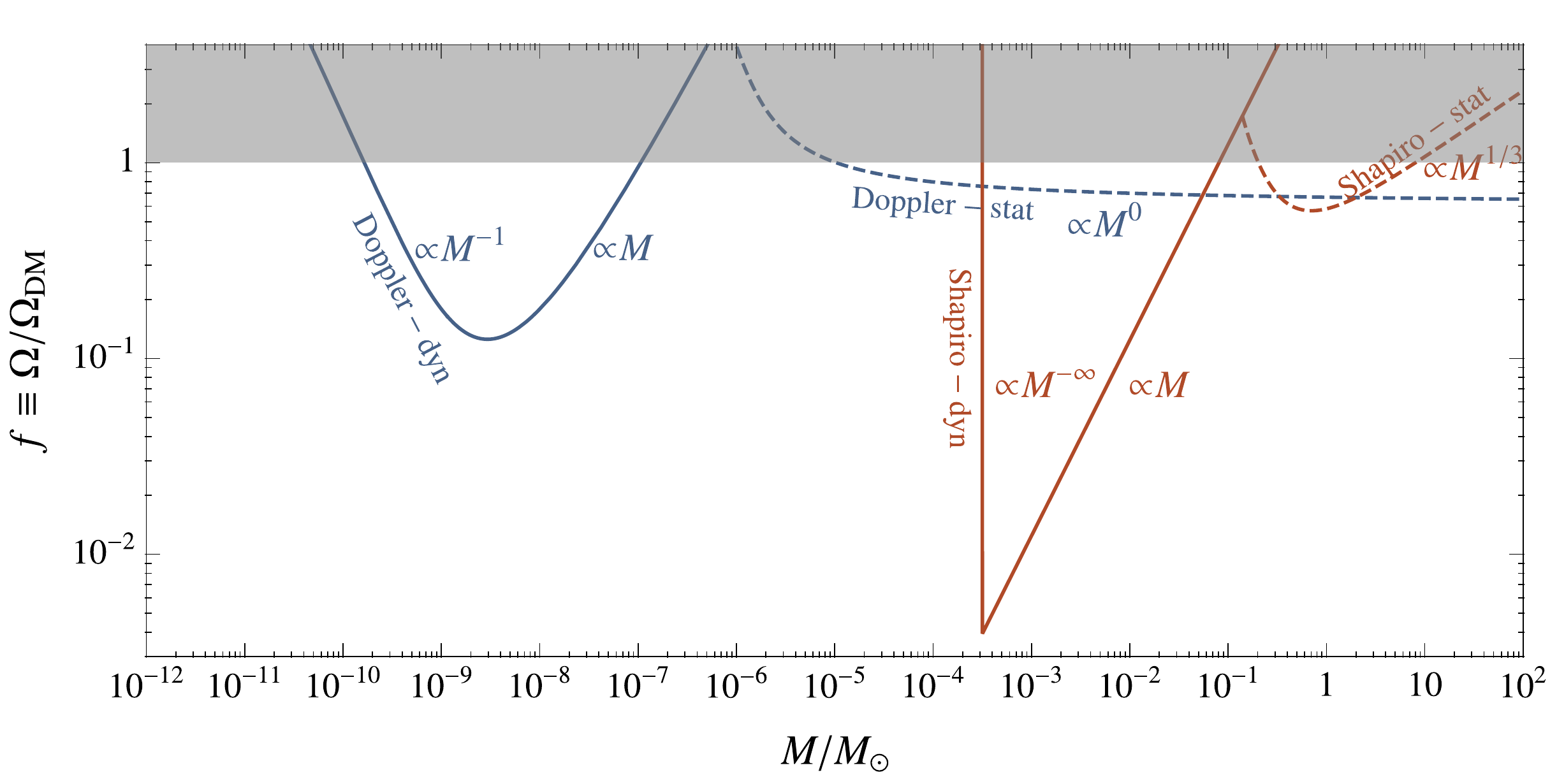}
\caption{SKA projected constraints on the DM fraction, $f$, for Doppler ({\color{c1} {\bf blue}}) and Shapiro ({\color{c2} {\bf red}}) signals contained in PBHs. The dynamic searches are shown in solid and static in dashed.  Each search is labeled with the mass scalings corresponding to the analytic formulae given in Eqs.~\eqref{eq:scalingdopdynamic}-\eqref{eq:scalingshstatic}.}
  \label{fig:MP2}
\end{figure*}

We now present our constraints on $f$ using the analytic formulae derived in the previous sections.  Note that the analytic formulae drop angular factors, assume a velocity, $v$, of $250$ km$/$s, and use the SKA PTA parameters given in Table~\ref{table:one}. Four of the subsequent constraint equations arise from equating the relevant SNR to four. The other two are simply reformulations of Eqs.~\eqref{eq:MStatDynDopp},~\eqref{eq:MStatDynShap}, which indicate the transition between the dynamic and static regions.  Our results are summarized in Fig.~\ref{fig:MP2}; these results are generated with a numerical simulation (detailed further below), but are consistent, to ${\cal O}(1)$ numbers, with the analytic results quoted in detail next.  The comparison between the analytic results and numerical simulation is discussed further (with a plot detailing differences) in Appendix~\ref{app:comparison}.

 \begin{figure*}[ht!]
   \includegraphics[width=0.75\textwidth]{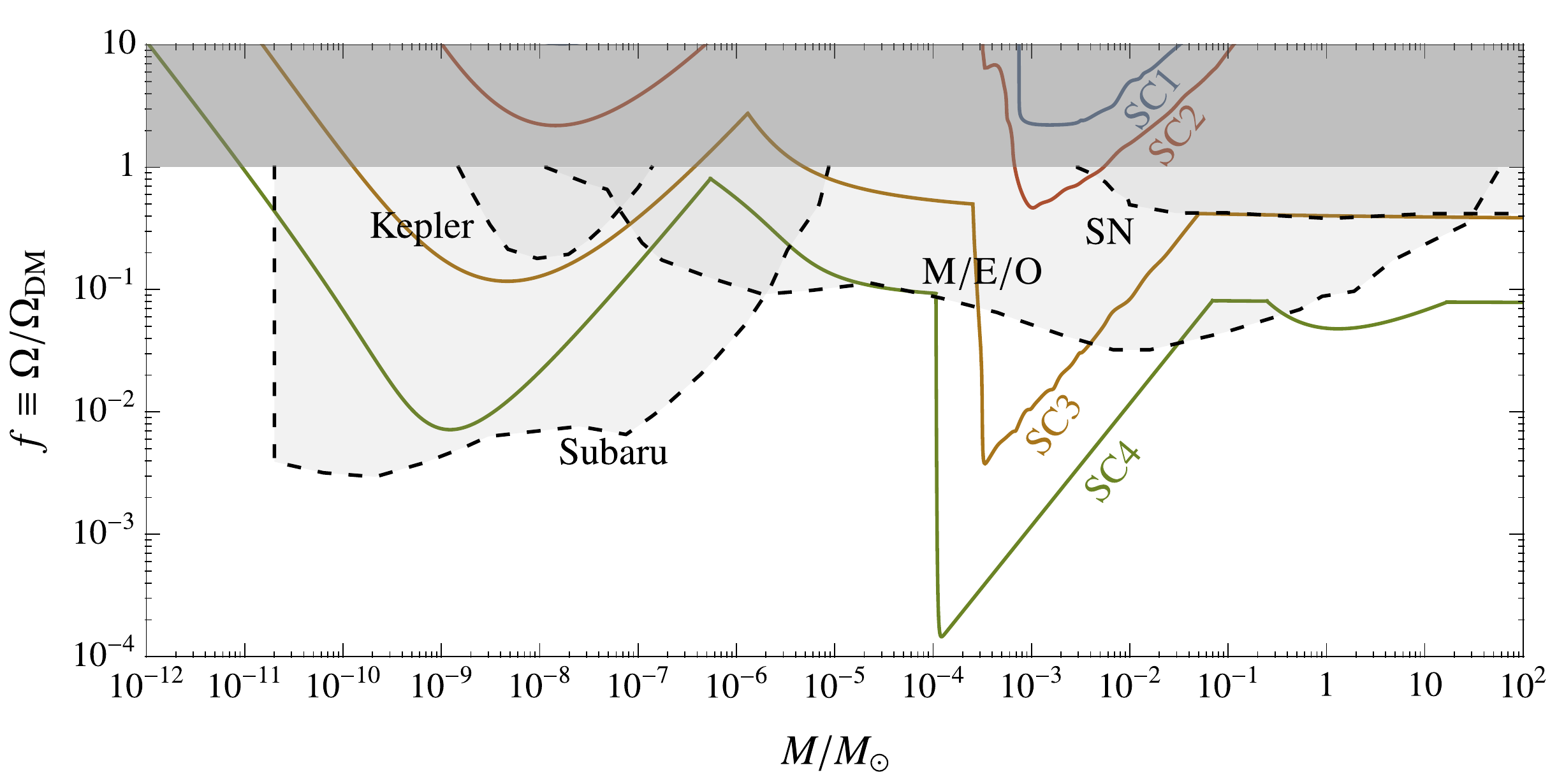}
 \caption{PTA projected constraints on PBH-like DM compact objects of mass $M$ and DM fraction, $f$, combining all proposed searches. The scenarios (labeled SC1-4) are described in the text and are roughly given by current capabilities ({\color{c1} {\bf blue}}), current capabilities with 10 more years of measurement on the same pulsars  ({\color{c2} {\bf red}}), SKA capabilities ({\color{c3} {\bf orange}}), and an optimistic scenario where SKA finds a large number of high performing pulsars ({\color{c4} {\bf green}}). The dark gray region corresponds to the unphysical case of $f>1$. For reference we also show the constraint from lensing (dashed {\color{black} {\bf black}}).}
   \label{fig:MP3}
\end{figure*}

To begin we consider the Doppler search in the dynamic limit. In this case dropping the angular factors in the SNR equation, Eq.~\eqref{eq:SNRdop}, equating the SNR to four, and substituting $\tau = \tau_\text{min}$ from Eq.~\eqref{eq:rminDopdynamic}, constrains $f$ to
\begin{align} 
f_{\text{D, dyn}}^L & \lesssim 0.1 \left( \frac{10^{-9} \, M_\odot}{M} \right) \left( \frac{200}{N_P}\right) \left( \frac{20 \text{ yr}}{T} \right)^4.
\label{eq:scalingdopdynamic}\end{align} 
The $L$ superscript denotes that this analytic constraint corresponds to the left-hand side of the triangular ``Doppler-dyn'' wedge in Fig.~\ref{fig:MP2}, labeled $ \propto M^{-1} $ at low masses. This behavior does not continue indefinitely, but is cut off when the closest object no longer satisfies our dynamic condition, $\tau_\text{min} \lesssim T/2 $, where $\tau_\text{min}$ is given by Eq.~\eqref{eq:rminDopdynamic}. This is equivalent to Eq.~\ref{eq:MStatDynDopp} and constrains $f$ to
\begin{align} 
f^R_{\text{D, dyn}} & \lesssim 3 \left(   \frac{ M }{  10 ^{ - 7}\, M _{\odot} }  \right) \left( \frac{ 200 }{  N_P}  \right) \left( \frac{20 \text{ yr}}{T} \right)^3   \,, 
\label{eq:scalingdopDynStat}
\end{align}
where the $R$ superscript indicates that this analytic constraint corresponds to the right hand side of the same triangular wedge in Fig.~\ref{fig:MP2}, labeled $\propto M$. 

We now repeat the above arguments for the other searches. The Shapiro constraint in the dynamic limit is obtained from equating the SNR, Eq.~\eqref{eq:SNRshap}, to four, which sets a lower bound on the masses reachable with a single event,
\begin{align}
M^L_\text{S, dyn} \approx 3 \times 10^{-4} \left( \frac{20 \, \text{yr}}{T} \right)^\frac{1}{2} \left( \frac{t_\text{RMS}}{50 \, \text{ns}}\right) M_\odot \label{eq:scalingshdynamic}
\end{align}
corresponding to the left-hand side of the wedge labeled ``Shapiro-dyn'' in Fig.~\ref{fig:MP2}.


Furthermore, the dynamic condition, $\tau_\text{min} < T/2$, with $\tau_\text{min}$ again given by Eq.~\eqref{eq:rminShdynamic} can be written, similar to Eq.~\ref{eq:MStatDynShap}, as a condition on $ f $ as,
\begin{align} 
f^R_\text{S, dyn} & \lesssim 0.8 \left( \frac{ M }{ 10 ^{ - 2 } \, M_\odot}\right) \left( \frac{ 200 }{ N_P }  \right) \left( \frac{20 \text{ yr}}{T}\right)^2 \,,
\label{eq:scalingshapDynStat}
\end{align} 
corresponding to the right-hand side of the Shapiro dynamic wedge in Fig.~\ref{fig:MP2} labeled $\propto M$.

Similarly analytic scalings in the static limit can also be derived. Equating Eq.~\eqref{eq:staticDop} to four and substituting $r_ 0 = r_\text{min}$ from Eq.~\eqref{eq:rminDopstatic} yields a constraint on $ f $ from the static Doppler search, 
\begin{align} 
f_\text{D, stat} & \lesssim 0.4 \left( \frac{200 }{ N_P } \right) \left( \frac{20 \text{ yr}}{T} \right)^\frac{7}{2}   \,.
\label{eq:scalingdopstatic}
\end{align} 
This corresponds to the curve labeled ``Doppler-stat'' and $\propto M^0$ in Fig.~\ref{fig:MP2}. Likewise equating Eq.~\eqref{eq:staticSh} to four and substituting $r_\times = r_{\times, \text{min}}$ from Eq.~\eqref{eq:rminShstatic} sets the constraint on $f$ from the static Shapiro search,
\begin{align}
f_\text{S, stat} & \lesssim  \left( \frac{200}{N_P}\right) \left( \frac{ M }{ M _{\odot} } \right)  ^{\frac{1}{3} }  \left( \frac{20 \text{ yr} }{T} \right) ^{\frac{7}{3} }  \left( \frac{ \text{kpc}}{ d } \right)  \,,
\label{eq:scalingshstatic}
\end{align}
which corresponds to the ``Shapiro-stat'' curve in Fig.~\ref{fig:MP2} labeled with scaling $\propto M ^{1/3} $. Note that, as mentioned earlier, the scaling of the static results at their low mass end, where the static approximation is breaking down, is not trivial since it does not follow the closest-object approximation. We do not attempt to study this in detail analytically but note that over most of the interesting parameter space, the static search in the highly static regime ($ \tau \gg T $) has the most promising reach. 

While the analytic results give the correct scaling and approximately correct magnitude, to set more rigorous constraints, which take into account the dark matter velocity distribution and all angular factors, we employ a Monte Carlo (MC) simulation which takes the PBHs to be monochromatic (of a single mass) and randomly distributed with density $ \rho _{ {\rm DM}} = 0.46 ~{\rm GeV} / {\rm cm} ^3 $~\cite{Sivertsson2017}, and with a Maxwell Boltzmann velocity distribution with a mean of $ 220~ {\rm km/s} $. The statistics employed are the same as discussed in the beginning of Sec.~\ref{sec:methods}. In both the dynamic and static limits we take the signal to be the largest SNR generated in the entire array of pulsars. 

We consider four different scenarios composed of the different pulsar sets. In the first scenario (SC1) we compute the constraints for only the current 73 pulsar set, assuming the search is done today. In scenario two (SC2) we assume the same set of pulsars, except the observation time of each is increased by ten years. In the third scenario (SC3) we consider the current pulsar set observed for 30 more years, and the addition of the set of pulsars from SKA measured for 20 years. This effectively assumes SKA will start taking data ten years from now with all current pulsars continuing to be studied. Lastly, scenario four (SC4) is the same as scenario three but with the optimistic parameters.  In this case the new pulsar measurements dominate the current pulsar measurements.

We present our full results in Fig.~\ref{fig:MP3}. The main peaks at $ 10 ^{ - 9 } - 10 ^{ - 8} \,M _{\odot}$ and $ 10 ^{ -4 } - 1 0 ^{ - 3} \,M _{\odot} $ arise from the dynamic Doppler and Shapiro searches. The non-smooth behavoir of the dynamic Shapiro curves is (largely) due to current pulsars each having different noise. While for other searches, the pulsar timing array will have an ``effective noise'' leading to smooth curves, the single blip approximation in Shapiro leads to a sharp cut in sensitivity for each pulsar in the set at a different mass (as a consequence of the $ \tau $-independence of the SNR) leading to the features shown in the figure. For SC3 and SC4, where the constraints dominantly arise from a new (assumed identical) set of pulsars from SKA, these features do not arise. At larger masses, the static searches become important with Doppler static becoming approximately constant at large masses and static Shapiro only becoming relevant for SC4 (where it gains due to its large assumed baseline). For comparison we include constraints from lensing experiments, which, as we will show in the next section, only apply for very compact objects. We see that even with the current set of pulsars, a dedicated search could begin to probe $ f < 1 $ within ten years. With the inclusion of additional pulsars from SKA, pulsar timing can scan a huge mass range, from as low as $ \sim 10 ^{ - 12} \,M _\odot $ and could constrain PBHs to sub-percent fractions of the DM. As already mentioned, in addition to the correct scalings, we also find that the MC and analytic results are in agreement up to $\mathcal{O}(1)$ factors, with the main differences being due to the angular factors, as well as all PBHs sharing a uniform velocity in the analytic approximation. 

Lastly, we note that there are several ways that these constraints could be drastically improved besides the addition of pulsars with better timing parameters, as can be seen in Eqs.~\eqref{eq:scalingdopdynamic},~\eqref{eq:scalingshdynamic},~\eqref{eq:scalingdopstatic}, and \eqref{eq:scalingshstatic}. For example, if a millisecond pulsar is found much farther away (say $ d \sim 10-100 \, \text{ kpc} $), the constraints from the Shapiro delay will improve, while a pulsar in a DM dense region (such as near the galactic center or in globular clusters) will yield stronger constraints for all proposed signals.

\section{Constraints on Subhalos}
\label{sec:size}

\begin{figure*} [ht]
  \begin{center} 
\includegraphics[width=\textwidth]{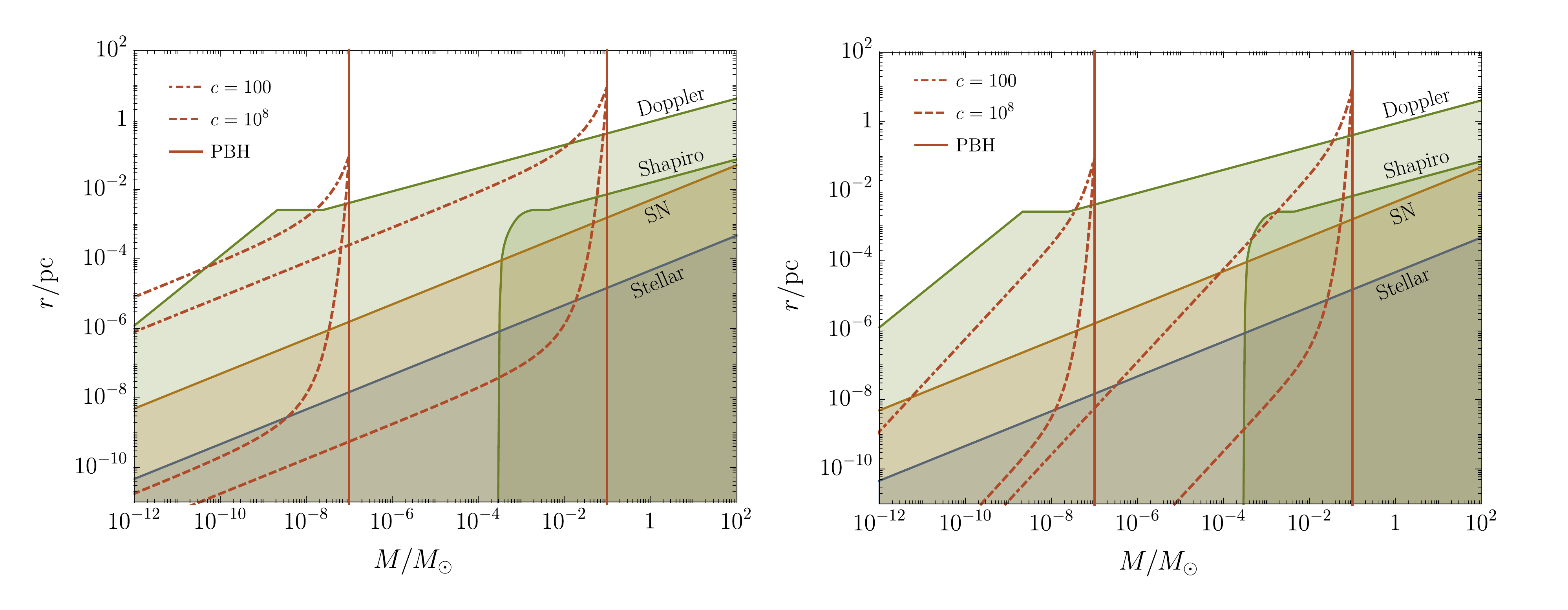} 
\end{center}
\caption{Sensitivity distance (as defined in the text) as a function of the compact object mass $M$ for PTA Doppler and Shapiro searches, as as well as supernova and stellar lensing.  The red curves in the left and right panels are the mass enclosed functions, $ r _{ {\rm enc}} $, for the NFW and UCMH-like profiles (defined in the text) respectively for three different concentration parameters $c$, and two different viral masses, $M_\text{vir} = 10^{-7} \, M_\odot, \, 10^{-1} \, M_\odot$.  Where these red curves intersect the sensitivity distance curves corresponds to the effective subhalo enclosed mass, $M^*$, to which the various searches are sensitive. See text for more details.}
\label{fig:size}
\end{figure*}

We now turn to constraining more diffuse DM subhalos, which as we now show, can be detected using the proposed searches for compact objects. In principle, diffuse subhalo signals can be calculated using the same procedure we invoked for compact objects, namely, computing $ \delta \nu / \nu $ for the Doppler and Shapiro signals, and finding the SNR using Eqs.~\eqref{eq:SNRstat} or \eqref{eq:SNR}, depending on the mass range of interest. The induced strain by a passing object is only dependent on the gradient of the gravitational potential due to the passing object (see Eqs.~\eqref{eq:dopsignalpre} and \eqref{eq:shapirosignalpre}). For a subhalo a distance $ r $ from the pulsar timing system, Gauss's law states that the gradient of the potential is given by,
\begin{align} 
\nabla \Phi (r)  & =  \frac{ GM  (r) \mathbf{r} }{ r^3 } \\
M (r) & = 4\pi \int_0^r  r'^2 \rho(r') \,  dr', \label{eq:halopot}
\end{align} 
where the integral runs from the center of the subhalo to the point of interest. Substituting this expression into~\eqref{eq:dopsignalpre} or~\eqref{eq:shapirosignalpre} gives $\delta \nu / \nu$ for a generic DM subhalo profile. Computing the signal shape and size is, however, rather involved; after all, $M ( r ) $ is time dependent as the halo moves. On the other hand, one can set conservative bounds by replacing $M$ in Eqs.~\eqref{eq:dnuDstatic}, \eqref{eq:dnuSstatic}, \eqref{eq:SNRdop}, and \eqref{eq:SNRshap}, by the minimum $M ( r ) $ over the time of observation, which is calculated at the point of closest approach of the halo to the pulsar (Doppler) or line of sight (Shapiro). 

The above lower bound is a particularly good estimate for dynamic signals. In this case the signal shape can be split up into two components, the one from the inner ring encapsulated by the impact parameter $b$ and an additional piece for the outer ring:
\begin{equation} 
\frac{ \delta \nu }{ \nu } =  \frac{ \delta \nu }{ \nu }\bigg|_{M(b) } + \frac{ \delta \nu }{ \nu } \bigg|_{ \text{ring}}\,.
\end{equation} 
A matched filtering prescription is optimized to look for a specific signal shape and hence would remove (part of) the additional contribution from the ring, leaving behind primarily a signal from the inner circle of radius $ b $. While more sophisticated analysis could improve and include these effects, it is beyond the scope of this work.

In the case of static signals, the non-compactness of the subhalo manifests itself as additional contributions in time derivatives of $M ( r ) $ in both the Doppler and Shapiro signals, as well a deviation in the integral over the line of sight, in the case of Shapiro. We have verified that these corrections contribute $\mathcal{O}(1)$ to the lower bound evaluated simply by taking $ M  ( r ) $ as evaluated at the initial position, $ M ( r _0  ) $.

To relate the limits computed earlier for PBHs to subhalos, it is convenient to define a {\em sensitivity distance}, which is the typical distance of a compact object to the pulsar (Doppler) or line of sight (Shapiro) in order to induce a particular SNR ({\em e.g.} four).  Limits on a particular DM fraction at a given mass are set when the minimum distance is smaller than this sensitivity distance.

The sensitivity distance in the case of the dynamic Doppler signal can be computed using Eq.~\eqref{eq:SNRdop}, and substituting SKA parameters given in Table~\ref{table:one} while taking $ v \sim 10^{-3}$, gives

\begin{align} 
r _{ {\rm PTA}}   & \sim  10^{-3} \text{ pc} \times  \left( \frac{M }{10^{-9} \, M_\odot}\right) \; \;  \text{(Doppler Dynamic)}.
\end{align} 
The Shapiro dynamic sensitivity distance is a more complicated function of mass, as the SNR is $\tau$ independent in the $\tau \ll T$ limit. The curve shown in Fig.~\ref{fig:size} is derived by taking the SNR given in Appendix~\ref{sec:simplifysignal} and finding the $r, M$ curve which satisfies $\text{SNR}(\tau = r/v, M) = 4$. 

\begin{figure*} [ht]
  \begin{center} 
\includegraphics[width=0.75\textwidth]{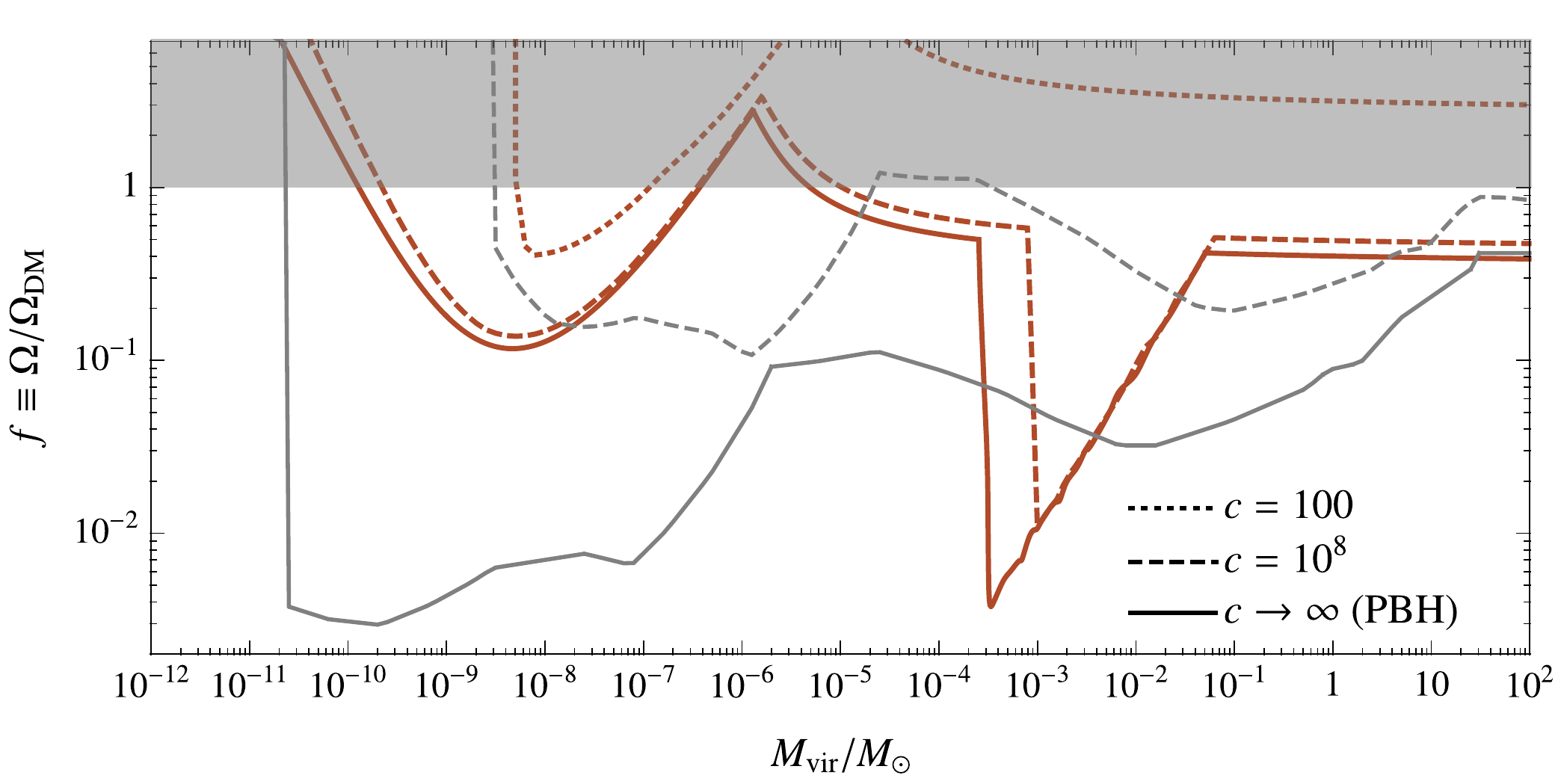} 
\end{center}
\caption{Comparison of the reach of pulsar timing data in scenario SC3 ({\color{c2} {\bf red}}) to lensing ({\color{gray} {\bf gray}}) experiments, for NFW halos of different concentration parameters, $c$.  With concentration parameters below $10^8$, PTA searches rapidly become more sensitive than lensing searches. The PTA constraints are cut-off when $ r_{ {\rm enc}} $ never crosses the sensitivity radius for any $ M $. For the concentration parameters chosen, this only occurs for the Doppler dynamic search leading to a sharp rise in the constraints at $M _{ {\rm vir}} = 10^{-9}~M_\odot$ for $ c = 100$.}
\label{fig:haloconstraints}
\end{figure*}

In the static limit, Eqs.~\eqref{eq:dnuDstatic} and \eqref{eq:dnuSstatic}  give the distances, %
\begin{align} 
r _{ {\rm PTA}}   & \sim  10^{-3} \text{ pc} \times \left\{ \begin{array}{lr} \left( \frac{M}{10^{-8} \, M_\odot} \right)^\frac{1}{3} & \text{(Doppler Static)} \\ \left( \frac{M}{10^{-3} \, M_\odot} \right)^\frac{1}{3}  & \text{(Shapiro Static)}\\ \end{array} \right. . \label{eq:sensitiver}
\end{align} 
We plot the sensitivity distances as a function of mass in Fig.~\ref{fig:size}. Note that in Fig.~\ref{fig:size} the dynamic curves end at $r = v T$ since, if the object passes at larger $r$, it would not satisfy the dynamic condition discussed earlier.

If the subhalo radius is smaller than the sensitivity distance then its effects on pulsar timing searches are identical to a PBH of the same mass. On the other hand, if the DM subhalo has a radius larger than the sensitivity distance, it may still be constrained if $r_\text{enc}$, the inversion of $M(r)$, is less than the sensitivity distance for some mass $M$. Physically, this means that the whole subhalo is too diffuse to measure but the core may be compact enough to measure.


To explore this possibility we consider two halo profiles of the generic form, 
\begin{align}
\rho(r, M_\text{vir}) & = \frac{\rho_s}{\left( r /r _s \right)^\alpha \left( 1 + r/ r _s \right)^\beta}\,, 
\end{align}
where $M_\text{vir}$ is the virial mass of the halo, $c \equiv r  _{ {\rm vir}} / r _s $ is the concentration parameter, $ r _{ \text{vir}} \equiv ( 3 M_\text{vir} / 800 \pi \rho _c ) ^{1/3} $ is the virial radius, and $\rho_s$ is an overall normalization factor fixed by requiring that the total mass inside of the virial radius is the virial mass. The standard NFW profile corresponds to taking $\alpha = 1, ~\beta = 2$, and an ultracompact minihalo~\cite{1984ApJ...281....1F,1985ApJS...58...39B} corresponds to $\alpha = 9/4, ~\beta = 0$ (though see, {\em e.g.},~\cite{Delos} which suggests an $ \alpha = 3/2 $, $ \beta = 3/2 $ profile can be a better fit to numerical simulations for halos produced from gravitational collapse of some primordial power spectra). 

In Fig.~\ref{fig:size} Left (Right) we plot $r_\text{enc}$ for NFW (UCMH-like $\alpha = 9/4, ~\beta = 3/4$) halos of virial mass $M_\text{vir} = 10^{-7} \, M_\odot$, $10^{-1} \, M_\odot$ and concentration parameters $c=100$, $10^8$ and the PBH-limit, $c \rightarrow \infty$. For a given subhalo, if $r_\text{enc}$ passes below a particular sensitivity curve at a mass, $M ^\ast $, then the search is sensitive to an effective subhalo of mass $M ^\ast $. These $M  ^\ast $ mass subhalos make up $f ^\ast =f \, M ^\ast/M_{\rm vir}$ of the DM. Hence the limits of interest, $( M_\text{vir} ,f )$, can be (conservatively) extracted from $( M ^\ast,f ^\ast )$, which is constrained for PBHs. Thus finally we obtain,
 \begin{equation}
 f =f ^\ast  \frac{M_\text{vir}}{M  ^\ast }=f_\text{PBH}(M ^\ast )\frac{M_\text{vir}}{M ^\ast } \,,\label{eq:rescale}
 \end{equation}
  where $ f _{ {\rm PBH}} ( M ^\ast ) $ it the limit extracted from Fig.~\ref{fig:MP3}. 
As we can see in Fig.~\ref{fig:size}, a Doppler search is sensitive to the largest radii, followed by a Shapiro delay search.

A similar procedure can be adopted to translate microlensing constraints for PBHs to subhalos. This time the sensitivity radius corresponds the Einstein radius, $ r _E $, above which microlensing experiments will not see modulations in the source brightness. The Einstein radius for a source and lens at distance $ D _S $ and $ D _L $ respectively is given by, 
\begin{equation} 
r _E \simeq \left( 4 GM \frac{ ( D _S - D _L ) D _L} { D _S  } \right) ^{1/2} \,.
\end{equation} 
The sources considered by microlensing experiments range from nearby stars~\cite{Griest:2011av,Griest:2013aaa}, the Large Magellanic Cloud (e.g.,\cite{Allsman:2000kg,Tisserand:2006zx}), Andromeda (e.g.,~\cite{Niikura:2017zjd}), and even distant supernova~\cite{Zumalacarregui:2017qqd}. For stellar sources, the distances are $\mathcal{O} (1 - 100 \text{ kpc}) $ while supernovae are sensitive to much larger distances, $ {\cal O} (  \text{Gpc} ) $. The typical  Einstein radii of these sources are,
\begin{align}
r_E  & \sim  \left\{ \begin{array}{lr} 10^{-6} \text{ pc} \left( \frac{M}{10^{-4} \, M_\odot}\right)^\frac{1}{2} & \text{(Stellar Lensing)}\\ 10^{-2} \text{ pc}\left( \frac{M}{10 \, M_\odot} \right)^\frac{1}{2} & \text{(Supernovae Lensing)} \end{array} \right. , 
\end{align}
and are illustrated in Fig.~\ref{fig:size}. For diffuse halos with radii much larger than this distance microlensing is unable to see a significant signal.  Note that because the Einstein radii are much smaller than the PTA  sensitivity distance, time delays due to the creation of multiple source images (as considered in Ref.~\cite{Katz:2018zrn}) are subdominant to the effects considered here.

Finally, we use Eq.~\eqref{eq:rescale} and the intersections in Fig.~\ref{fig:size} to find limits on the dark matter fraction $f$ as a function of the total DM subhalo mass, $M _{ {\rm vir}} $, in Fig.~\ref{fig:haloconstraints} for the SKA pulsar set defined in Sec.~\ref{sec:results} as SC3. The lensing constraints are scaled in the same way as pulsar timing, where we take the sensitivity line to be the characteristic Einstein radius from objects in the Andromeda galaxy for the Subaru constraint and from objects in the Milky Way for the MACHO/Eros/Ogle constraint curve.

We observe in Fig.~\ref{fig:haloconstraints} that in the PBH-limit, in most of the mass range, constraints in SC3 are stronger for lensing compared to pulsar timing; however as the concentration parameter is decreased, lensing drops off in sensitivity relative to PTAs. We find that for NFW halos with $c = 10^8$ the Subaru search is only sensitive to halos of $ M \gtrsim 10^{-9}~M_\odot$. For a CDM-inspired \cite{Sanchez-Conde:2013yxa} concentration parameter $c = 100$, we no longer observe any sensitivity from lensing, while PTAs can constrain a non-negligible $f$.  Note, however, that this sensitivity occurs only for very low halo masses (significantly below an Earth mass) where one expects halo disruption.  On the other hand, for slightly higher concentration parameters $c = 10^3$, sensitivity to $f < 1$ occurs similarly to $c = 100$ for low mass halos, but also (from a static Doppler search) for $M \gtrsim 10^{-3} \, M_\odot$. 

While our analysis of diffuse halos is schematic and suffers from $ {\cal O} ( 1 ) $ corrections, it serves to emphasize the complementarity between lensing and pulsar timing probes. Fully exploiting the potential of PTAs to constrain diffuse halos and specific models of structure formation is an intriguing problem which we leave for future work~\cite{future}.

\section{Conclusions}
\label{sec:conclusion}
In this work we considered pulsar timing constraints on DM compact objects, focusing on primordial black holes and subhalos. We studied pulsar timing signatures over the mass range from $10^{-12} \,M_\odot$ to $100 \,M_\odot$ finding that, depending on an objects mass, different searches are required to detect it. We examined four different types of searches, dynamic and static signals, each arising from Doppler and Shapiro time delays. Importantly, we computed the signals in three-dimensions and highlighted their relation to one another. Furthermore, using a Monte Carlo analysis we performed projections for pulsar timing capabilities using current and future pulsar timing experiments and understood their scaling using analytic techniques. With dedicated searches we found that current pulsar timing arrays can, over the next decade, set non-negligible constraints through dynamic searches.  Farther into the future, we expect sub-percent level constraints over the entire range with upcoming pulsar timing arrays.
 
There are two primary ways in which the capabilities of detecting DM compact objects from pulsar timing can be drastically improved. First, we assumed that all DM compact objects are in regions with DM density comparable to our local density. If instead pulsars are discovered in DM-dense regions ({\em e.g.}, close to the galactic center or within dwarf galaxies), then it is possible to quickly improve the power of Doppler signals. Similarly if pulsars are discovered with a line of sight passing through a DM-dense region, then the capabilities of a Shapiro search will be greatly enhanced. Second, the Shapiro search potential is sensitive to the distance to the pulsar (in the case of uniform density, limits on the fraction of the DM constrained scale linearly with the baseline), such that if pulsars are discovered significantly farther from our local neighborhood ({\em e.g.}, extra-galactic pulsars), then the Shapiro search will quickly become more powerful.

The constraints studied here apply to substructure which has survived to the present day, and we do not attempt to relate these substructures to specific astrophysical or particle physics models. Relating structure on such small scales to a model is a challenging exercise due to the uncertainties on the survival of low-mass subhalos to the present epoch.  Previously, Refs.~\cite{Baghram:2011is,Kashiyama:2018gsh} considered evolution of subhalos in a vanilla Cold DM (CDM) paradigm with Stable Clustering and spherical halo models to predict halo substructure, and came to opposite conclusions about the feasibility of detecting DM substructure in the CDM model with PTAs.  The difference in the conclusions of these papers is likely partly due to the DM clustering model and partly due to a difference of methods with respect PTA constraints.  Utilizing the methods proposed in this paper, we plan to consider PTA constraints on vanilla CDM, and other models such as axion miniclusters, in future work~\cite{future}.  If theory predictions can be made reliably, pulsar timing will become a powerful tool to probe the nature of DM.

Finally we note that our analysis was entirely focused on single events. For lower masses (below $ \sim 10 ^{ - 9} \,M _{\odot} $ for the Doppler dynamic search and below $ \sim 10 ^{ - 4} \,M _{\odot} $ for the Shapiro dynamic search), we expect multi-event signals that are not detectable as single events to become important. Nevertheless, they may leave an imprint in pulsar timing arrays that can detected using a statistical prescription as considered in~\cite{Baghram:2011is}. We leave further study of this limit to future work~\cite{future}. Taken together, however, a coherent picture is emerging for how compact objects can be constrained across a huge mass range with one observational tool.

\section*{Acknowledgments:}
We thank Adrienne Erickcek, Andrey Katz, Joachim Kopp, Vikram Ravi, Katelin Schutz, Sergei Sibiryakov, Wei Xue, Xingjiang Zhu, and Miguel Zumalacarregui for useful discussions. We especially thank Stephen Taylor for extensive discussions and collaboration during early parts of this work. JD, HR, TT and KZ are supported in part by the DOE under contract DE-AC02-05CH11231. KZ thanks the CERN theory group for hospitality for the duration of this work.  Some of this work was also completed at KITP, supported in part by NSF Grant No. NSF PHY-1748958, and at the Aspen Center for Physics, which is supported by NSF grant PHY-1607611.

\appendix

\section{Simplification of Doppler and Shapiro signals and SNR}
\label{sec:simplifysignal}

This Appendix contains many of the details for the results presented in Sec.~\ref{sec:ShapiroVsDoppler}, on the signal shapes for the Doppler and Shapiro delays. In both cases, we work in the limit where the trajectory of the compact object is unaffected by the presence of the pulsar-Earth system (this amounts to assuming highly unbound orbits for the Doppler signal and a large baseline for the Shapiro signal). In either case we define a ``time until blip center'', $ t _0 $,  and a ``signal width'', $ \tau $, corresponding to the dot-product and cross-product of the distance-velocity vectors respectively.
\subsection{Doppler}

\subsubsection{Signal}
We begin by considering the Doppler effect, where our goal is to compute the velocity of the pulsar as a function of time,
\begin{equation} 
  {\mathbf{v}} _P = GM \int  \frac{  {\bf r}   }{ r ^3  } \, dt,
\end{equation} 
where $ {\bf r} $ is the vector pointing from the pulsar to the compact object and is taken to be unaffected by the presence of the pulsar, $ {\bf r} = {\bf r} _0 + {\mathbf{v}} t $. It is convenient to introduce the time variables,
\begin{equation} 
t _0 \equiv - \frac{ {\bf r} _0 \cdot {\mathbf{v}} }{ v ^2 } \quad; \quad  \tau \equiv \frac{ \left| {\bf r} _0 \times {\mathbf{v}} \right| }{ v ^2 } 
\end{equation} 
such that $ t _0 ^2 + \tau ^2 = r _0 ^2 / v ^2 $. In this case the magnitude of the position is
\begin{align} 
r & = \sqrt{r _0 ^2 - v ^2 t _0 ^2 + v ^2 ( t - t _0 ) ^2} \nonumber \\ 
& = v \tau ( 1 + x ^2 ) ^{1/2} ,
\end{align} 
where we have defined a normalized time, $ x \equiv ( t -  t _0 ) / \tau $. This gives a resultant strain,
\begin{align} 
\frac{ \delta \nu  }{ \nu} & = \frac{ GM }{ v ^3 \tau ^2  } \int d x \, \frac{1}{ ( 1 + x ^2 ) ^{3/2} } \left[ {\bf r} _0 + {\mathbf{v}} ( \tau x + t _0 ) \right] \cdot {\bf \hat{d} } \nonumber \\ 
                           & = \frac{ G M }{ v ^2 \tau } \frac{1}{ ( 1 + x ^2 ) ^{1/2} } \left( x {\bf \hat{b} } - \mathbf{\hat{v}} \right) \cdot {\bf \hat{d} } , \label{App:Doppdelnu}
\end{align} 
where we defined the impact parameter vector $ {\bf b} \equiv  {\bf r} _0 + t _0 {\mathbf{v}} $, obeying $ \left| {\mathbf{b}} \right| = v \tau $.  We thus obtain the result in Eq.~\eqref{eq:dopplersignalfinal}.

\subsubsection{SNR}
Here we derive Eq.~\eqref{eq:SNRdop} from the signal Eq.~\ref{App:Doppdelnu} in the $\tau, \left| t_0 \right|  \ll T$ limit. The signal must be windowed in order to account for the finite time of the experiment and we do this with a top-hat function in the time domain. The SNR from Eq.~\eqref{eq:SNR} in the $ \left| t_0 \right|  \ll T$ limit then becomes
\begin{align}
\text{SNR}^2 =  & \frac{G^2 M^2}{2 \pi^2 v^4t_\text{RMS}^2 \Delta t} ( {\bf \hat{b} } \cdot {\bf \hat{d} } ) ^2 \int_0^\infty \int_{-T / 2 \tau }^{T/ 2 \tau } \int_{-T / 2 \tau }^{T / 2 \tau }df \, dx \, dy \nonumber \\
& \times \frac{e^{2\pi i f \tau (x-y)}}{f^2} \frac{xy}{\sqrt{(1+x^2)(1+y^2)}} \label{eq:SNRDopAppen}
\end{align}
the frequency integral can be regulated with a fictitious ``mass'' term in the denomator parameterized by $\epsilon $
\begin{align}
\int_0^\infty df \, \frac{e^{2 \pi i f \tau(x-y)}}{f^2 + \epsilon^2} & = -\pi^2 \tau |x-y| + ... \label{eq:regulateF}
\end{align}
where the ellipses denote terms which are odd in $ x $ or $ y $ and therefore vanish when the $ x,y $ integrals are taken (including the divergent piece). Inserting this into the above and carrying out the $x,y $ integrals we find the $ \tau \ll T  $ limit,
\begin{equation} 
 {\rm SNR}   \simeq  \frac{ G M }{ \tau v ^2 }  \sqrt{  \frac{ T ^3 }{ 12 \, t _{ {\rm RMS}} ^2  \Delta t }} \,{\bf \hat{b} } \cdot {\bf \hat{d} }\,,
\end{equation} 
in agreement with Eq.~\eqref{eq:SNRdop}.

\subsection{Shapiro}

\subsubsection{Signal}
We now compute the signal shape in the case of a Shapiro time delay starting with~\eqref{eq:shapirosignalpre}. The gradient of the PBH potential, $\Phi$, is evaluated along the line of sight (taken in the $ {\bf \hat{z} } $ direction), giving
\begin{equation} 
\frac{ \delta \nu }{ \nu  } = - 2 G M \int  \frac{ \dot{r} }{ r^2 } \, dz.
\end{equation} 
It is convenient to use coordinates that make the cylindrical symmetry of the problem manifest, $ {\bf r} _\times \equiv {\bf r} \times {\bf \hat{d} }$ and $ {\mathbf{v}} _\times \equiv {\mathbf{v}} \times {\bf \hat{d} } $, which results in time variables with the same interpretations as for Doppler:
\begin{equation} 
t _0 \equiv - \frac{ {\bf r} _{ \times , 0} \cdot  {\mathbf{v}} _\times }{ v _\times ^2 }\quad ; \quad \tau \equiv  \frac{ \left| {\bf r} _{ \times , 0} \times {\mathbf{v}} _{ \times } \right| }{ v _\times ^2 } .
\end{equation} 
Carrying out the integral for signals close to the line of sight we can rewrite the strain in terms of these coordinates as
\begin{equation} 
  \frac{ \delta \nu }{ \nu }= - 4 GM \frac{ \dot{r} _\times }{ r _\times}\,. \label{eq:ssstrain}
\end{equation} 
As in the Doppler signal we can rewrite the magnitude of the signal shape as
\begin{align} 
r _\times & = \sqrt{r _{ 0, \times } ^2 - v _\times  ^2 t _0 ^2 + v _\times ^2 ( t - t _0 ) ^2} \\ 
& =  v _\times \tau ( 1 + x ^2 ) ^{1/2} .
\end{align} 
We have again defined $ x \equiv ( t - t _0 ) / \tau $. Inserting this into~\eqref{eq:ssstrain} gives the signal shape
\begin{equation} 
\frac{ \delta \nu }{ \nu } = - \frac{ 4 GM }{ \tau } \frac{ x }{ 1 + x ^2 }. \label{App:Shapdelnu}
\end{equation} 

\subsubsection{SNR}
Here we derive Eq.~\eqref{eq:SNRshap} from Eq.~\ref{App:Shapdelnu} in the $\left| t_0 \right| , \tau \ll T$ limit. Analogous to our approach in deriving the Doppler SNR, in the time domain we window our function for a finite observing time, giving
\begin{align}
\text{SNR}^2 = & \frac{8 G^2 M^2}{\pi^2 t_\text{RMS}^2 \Delta t} \int_0^\infty \int_{-T / 2 \tau }^{T / 2 \tau } \int_{-T / 2 \tau }^{T / 2 \tau } df \, dx \, dy \nonumber \\
& \times \frac{e^{2\pi i f \tau (x-y)}}{f^2} \frac{xy}{(1+x^2)(1+y^2)} . \label{eq:SNRShAppen}
\end{align}
As for the Doppler signal we regulate using~\ref{eq:regulateF}. In the $\tau \ll T$ limit the SNR becomes 
\begin{equation} 
 {\rm SNR}    \simeq   G M \sqrt{  \frac{ 32 \, T }{ t _{ {\rm RMS}} ^2 \Delta t } }\,,
\end{equation} 
in agreement with Eq.~\eqref{eq:SNRshap}.

\section{Minimum Distance Statistics}
\label{sec:mindist}

This Appendix is dedicated to deriving the minimum distances and times quoted in Sec.~\ref{sec:methods}.  Each of the signals considered (Doppler/Shapiro, dynamic/static) depends on the relevant distance of objects described by a random variable, $B$. The dominant signal comes from the object which has the minimal $B$, and therefore to gain an analytic understanding of the dominant signal we calculate the $100 \times p$\textsuperscript{th} percentile of $B_\text{min} \equiv \text{min} \left( B_1, B_2, ..., B_N \right)$. Every statistic of $B_\text{min}$ can be calculated with its cumulative distribution function (CDF), $F_{B_\text{min}}$, and because the $B_i$'s are independent and identically distributed random variables we can solve for $F_{B_\text{min}}$ solely in terms of $F_B$:  
\begin{align}
    F_{B_\text{min}}(b) & \equiv 1 - \text{Pr}(B_\text{min}  \geq b) \nonumber \\
    & = 1 - \text{Pr}(B_1 \geq b \cap B_2 \geq b \cap ...) \nonumber \\
    & = 1 - \left( \text{Pr}\left( B_1 \geq b \right) \right)^N =1 -  (1 - F_B(b))^N. \label{eq:mincdf}
\end{align}
The $100 \times p$\textsuperscript{th} percentile of $B_\text{min}$ is calculated by solving $F_{B_\text{min}}(b_p) = p$. For each signal we take $N$ objects populated inside the relevant volume. At the end of this Appendix we show the comparison between calculating the constraints on $f$ using the distances derived here and our Monte Carlo simulation.

\subsection{Doppler}

\subsubsection{Static}
For this signal the relevant distance is simply the distance from the object to a pulsar (we take the pulsar to define the origin). Assuming the objects are uniformly distributed in a sphere of radius one means $F_B(b) = b^3$. Therefore in the large $N$ limit,
\begin{align}
    b_p &= \left( - \frac{\ln \left( 1 - p \right)}{N} \right)^\frac{1}{3} \, .
\end{align}
For a sphere of radius $R$, we multiply $b_p$ by $R$ and take $N = 4 \pi N_P n R^3 /3$, giving 
\begin{equation}
b_p = - \left( \frac{3}{4 \pi}  \frac{\ln{\left( 1-p \right)}}{N_P n}\right)^\frac{1}{3}.
\end{equation}
Substituting $p = 0.9$ gives \eqref{eq:rminDopstatic}.

\subsubsection{Dynamic}

Here the relevant distance is the impact parameter of the passing object. Being only concerned with an order of magnitude estimate, we simplify the problem of finding the CDF of $B$ by taking the objects to move in the same direction. The impact parameter is then uniformly distributed across the cross sectional area. Therefore inside of a circle of radius one, in the cross section centered at the pulsar, the CDF of $B$ is $F_B = b^2$ and therefore in the large $N$ limit,
\begin{align}
    b_p & = \sqrt{-\frac{\ln \left( 1-p \right)}{N}}  \label{eq:bpDopdyn}
\end{align}
To find $N$ we calculate how many objects are in the dynamic limit. The dynamic limit is defined by $T - \tau > t_0 > \tau$. Geometrically this constrains how far an object can be as a function of it's impact parameter and defines a conical volume. Therefore in time $T$, $N = N_P n \pi v T R^2 / 3$. Multiplying \eqref{eq:bpDopdyn} by $R$ and substituting in $N$ gives
\begin{equation}
b_p = \sqrt{- \frac{3}{\pi}\frac{\ln{\left( 1- p \right)}}{N_P n v T}}.
\end{equation}
Substituting $p = 0.9$ and dividing by $v$ gives Eq.~\eqref{eq:rminDopdynamic}.

\subsection{Shapiro}
\subsubsection{Static}
The relevant distance here is the distance from a PBH to the central axis of a cylinder. Assuming the objects are uniformly distributed in a cylinder of radius one, and length $d$, means $F_B(b) = b^2$. Therefore in the large $N$ limit,

\begin{align}
    b_p & = \sqrt{- \frac{\ln\left( 1 - p \right)}{N} }
\end{align}
Taking $N = N_P n \pi d R^2$, while multiplying by $R$ to for a cylinder of radius $R$, gives 
\begin{equation}
b_p = \sqrt{\frac{-\ln{(1 - p)}}{\pi} \frac{1}{N_P n d} }.
\end{equation}
Taking $p = 0.9$ gives \eqref{eq:rminShstatic}.
\begin{figure*} [ht]
  \begin{center} 
\includegraphics[width=0.75\textwidth]{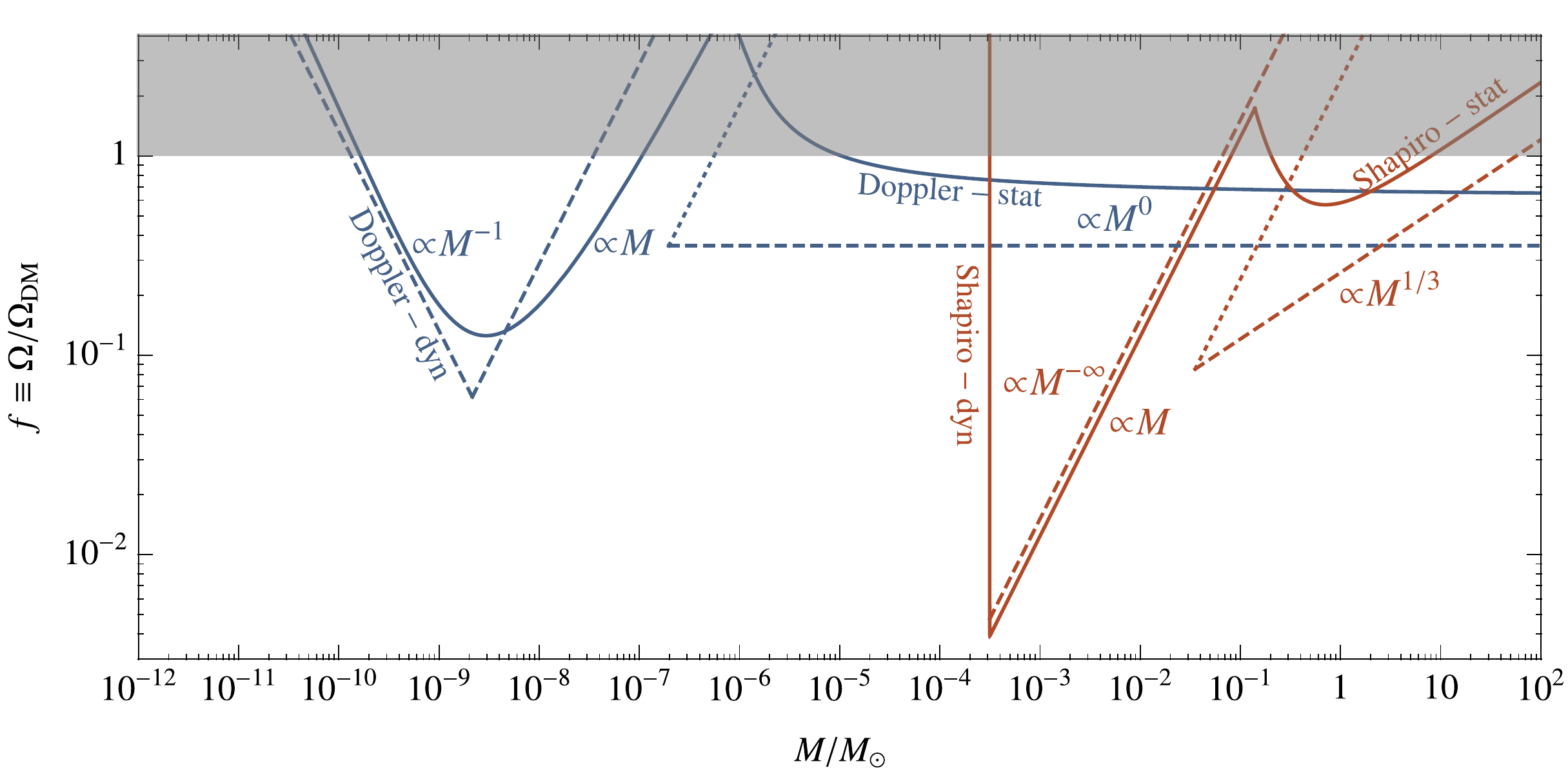} 
\end{center}
\caption{Comparison of the analytic estimates (solid) and the full Monte Carlo approach (dashed) for SKA PTA parameters given in Table~\ref{table:one}. The analytic estimate agree with the numerical results to $ {\cal O} ( 1) $ corrections except on the left-hand side of the static constraints, where the closest-object approximation breaks down.}
\label{fig:appendixcompare}
\end{figure*}

\subsubsection{Dynamic}
Analogous to the Doppler dynamic signal, the relevant distance here is the impact parameter defined from the central axis of a cylinder. Again, being only concerned with an order of magnitude estimate, we simplify the problem of finding the CDF of $B$ by taking the objects to move in the same direction. The impact parameter is then uniformly distributed across the cross sectional area. Therefore inside a rectangle of width $d$ and height one, the CDF of $B$ is $F_B = 2 b$ such that in the large $N$ limit, 
\begin{equation}
	b_p = -\frac{\ln (1-p)}{2 N} \label{eq:bpShapdyn}
\end{equation}
The dynamic limit is defined by $T - \tau > t_0 > \tau$. Geometrically this constrains how far an object can be as a function of its impact parameter, defining a parallelogram. Therefore in time $T$, $N = N_P n v T d D / 2$. Multiplying \eqref{eq:bpShapdyn} by $D$, for a cylinder of diameter $D$, and substituting $N$ gives 
\begin{equation}
b_p = -\frac{\ln{(1-p)}}{N_P n v T d}.
\end{equation}
Taking $p = 0.9$ and dividing by $v$ gives Eq.~\eqref{eq:rminShdynamic}.


\section{Comparison of Analytic and Monte Carlo Constraints}
\label{app:comparison}
We now make an explicit comparison in Fig.~\ref{fig:appendixcompare} between the analytic and numeric results presented in Sec.~\ref{sec:results}.  Six of the eight analytic curves are described by Eqs.~\ref{eq:scalingdopdynamic},~\ref{eq:scalingshdynamic},~\ref{eq:scalingdopstatic}, and ~\ref{eq:scalingshstatic} for the four signal types, and by Eqs.~\ref{eq:scalingdopDynStat} and \ref{eq:scalingshapDynStat} which are the boundary of the dynamic regimes. The other two (dotted) curves are $r_{\text{min}} = v T$, where $r_{\text{min}}$ is found in Eq.~\eqref{eq:rminDopstatic}, and $r_{\times, \text{min}} = v T$, where $r_{\times, \text{min}}$ is found in Eq.~\eqref{eq:rminShstatic}. At first glance these lines seem to be in disagreement with the Monte Carlo. However one must remember that the these only indicates when the \textit{closest} object is in the static limit, and the Monte Carlo is summing the contributions to $\ddot{\nu}$ of every object. Therefore this left hand side tells us where we would should expect the Monte Carlo and analytic methods to start diverging, consistent with what is observed. 

As can be seen in Fig.~\ref{fig:appendixcompare}, at least deep in the static or dynamic regimes where analytic approximations are expected to hold, the analytic estimates differ from the full Monte Carlo results by $\mathcal{O}(1)$ factors in the scaling regimes. This is due to the two main differences in the analytic and Monte Carlo results. The first difference is that in the full Monte Carlo the velocity of each object is drawn from a Maxwell Boltzmann distribution whereas the analytic results assume $v = 250$ km/s. The second difference is because analytic results drop the angular dependence of some expressions, such as $\mathbf{\hat{b}} \cdot \mathbf{\hat{d}}$ in Eq.~\eqref{eq:SNRdop}.

\bibliography{PTA4DM}

\end{document}